\documentclass{apa7}
\usepackage[american]{babel}
\usepackage{pslatex}
\usepackage{float}
\usepackage{csquotes}
\usepackage{microtype}
\usepackage{amsmath}
\usepackage{amssymb}
\usepackage{amstext}
\usepackage{multirow}
\usepackage{makecell}
\usepackage{tabularx}
\usepackage{graphicx}
\usepackage[table]{xcolor}
\usepackage{tikz}

\newcommand{\splitcolor}[3]{%
  \tikz[baseline=(base.base)]{
    \node[inner sep=0pt, outer sep=0pt, text opacity=0] (base) {#3};

    \begin{scope}
      \clip (base.south west |- base.west) rectangle (base.north east);
      \node[inner sep=0pt, outer sep=0pt, anchor=base] at (base.base)
        {\textcolor{#1}{#3}};
    \end{scope}

    \begin{scope}
      \clip (base.south west) rectangle (base.north east |- base.west);
      \node[inner sep=0pt, outer sep=0pt, anchor=base] at (base.base)
        {\textcolor{#2}{#3}};
    \end{scope}
  }%
}
\graphicspath{{./figures/}}
\usepackage[backend=biber, style=apa, natbib=true, sorting=nyt, sortcites=true]{biblatex}
\usepackage{software-biblatex}
\DeclareLanguageMapping{american}{american-apa}
\title{Rate-Coding Bundle Memory: A Unified Model of Memory and Control for Symbolic Computation in the Brain}
\shorttitle{Rate-Coding Bundle Memory}
\leftheader{Rate-Coding Bundle Memory}

\authorsnames[1,1,1,{1,2}]{Teun van Gils, Rowan P. Sommers, Markus Ostarek, Peter Hagoort}
\authorsaffiliations{{Neurobiology of Language Department, Max Planck Institute for Psycholinguistics}, {Donders Institute for Brain, Cognition and Behaviour, Radboud University}}

\authornote{
	Correspondence concerning this article should be addressed to Teun van Gils, Neurobiology of Language Department, Max Planck Institute for Psycholinguistics, Wundtlaan 1, 6525 XD, Nijmegen, The Netherlands.  
	E-mail: Teun.vanGils@mpi.nl
}

\abstract{
	We propose a neurobiologically plausible model of cognition that combines the advantages of connectionist and symbolic systems, and that can explain a wide range of cognitive phenomena. 
	This model, called Rate-Coding Bundle Memory (RCBM), is based on the Symbolic Subsystem Hypothesis, which posits that the brain implements a symbolic subsystem within its fundamentally connectionist nature.
	RCBM is a hybrid model that uses rate coding to represent symbols in a continuous space, and it uses a bundle memory system to store and retrieve these symbols.
	The model is capable of solving a wide range of cognitive phenomena, including one-shot learning, pattern separation, and the binding problem.
	We argue that RCBM provides a promising framework for understanding the nature of cognition, and that it can be used to develop more sophisticated models of cognition in the future.
}

\begin{document}
	\maketitle

	\section{Introduction}\label{sec:introduction}

	Few things defy a simple explanation as much as our mental model of the world: for any categorical, any box we create, there's an edge case, a nuance, a contradiction to be found.
	We may not usually pay much attention to the concepts we use, but when we do, we often find that they are unexpectedly tricky to pin down.
	Take a concept like ``bird'': it's easy to think of a robin or a sparrow, but what about a penguin, or an ostrich?
	What about a chicken, or its ancient ancestor, the T-Rex?
	Or what about a bat, or a dragonfly?
	Even though we can categorize these animals as birds or not, the boundaries of the category are fuzzy, and when asked to provide a rigid definition, we might struggle to find one that both conforms to our intuitions and covers all cases \citep{wittgenstein_philosophical_2010,rosch_natural_1973}.
	Is it being able to fly? Is it having feathers? Is it being warm-blooded? Is it having a beak? Is it being able to lay eggs?
	The answer is, perhaps, that it's all of these things and none of them.
	Our categorizations do not exist in a vacuum, but are always situated in a particular context and serve a particular goal, a position known as goal-directedness in embodied cognition \citep{barsalou_abstraction_2003,barsalou_grounded_2008}.
	Within this context, we can see that the concept of ``bird'' is not a single, fixed entity, but a complex, dynamic network of associations activated in parallel, with a grounding in the physical world and how we interact with it, and with meaning that is not fixed, but rather lies on a continuum of related concepts, each with their own fuzzy boundaries \citep{barsalou_grounded_2008,varela_embodied_2016}.
	A connectionist system, one comprised of distributed activations over simple, interconnected units that can learn input-output associations adaptive for their body-environment interactions is well-suited to model this kind of graded and continuous meaning \citep{hinton_distributed_1984,rogers_parallel_2014,doerig_neuroconnectionist_2023}.
	It provides a way to reason about the world that is robust against noise and small differences, and that allows for associative reasoning and graded (i.e., less rigidly categorical) inference \citep{hinton_distributed_1984,rogers_parallel_2014}.
	Though there are many different flavors of the aforementioned (connectionist) models and theories, they generally share a semantic system where information is distributed over different units in the network, where meaning is determined by associations between these units and their embodied relations to the world, and where the activation of these units is graded and continuous and can happen in parallel across the network.
	In other words, they pose that the brain is PACED: Parallel, Associative, Continuous, Embodied and Distributed.
	Given their success in psychology, neuroscience and more recently in AI, do we need anything else than connectionist systems to explain cognition?
	
	Imagine reading a novel or watching a movie where a plethora of different characters are introduced and interact in a complex plot.
	After finishing the story most people will not be able to remember the fine-grained details about each specific chapter or scene, such as the exact words used or the visual details of objects in the background.
	Yet, we are able to recall most of the different (relevant) characters, objects, locations, their attributes, and their relations to one another.
	Intuitively, we seem to store these different entities and their attributes as distinct items in memory, which we can then retrieve and recombine to form a mental model of the story.
	To explain this, it has been hypothesized that we possess a symbolic system to achieve this feat \citep{fodor_language_1975,marcus_algebraic_2001,frankland_language_2019}, allowing us to form semantically arbitrary combinations of discrete symbols.
	Symbols are semantically arbitrary in two ways: first, symbols are semantically detached from the worldly entities that they refer to, and thus any symbol could in principle refer to any entity, as long as there is a shared agreement on this reference.
	For instance, the word ``bird'' could just as well refer to a cat, and the word ``cat'' could just as well refer to a bird -- it would take some getting used to, but there is no inherent reason why this would not be possible.
	After all, the word ``bird'', as a collection of letters, bears no direct relation to or resemblance with the actual animal.
	More broadly, symbols do not need to be words; \citet{fodor_language_1975} argues that the mental representations that we use to think and reason about concepts are themselves symbols, forming what he calls a \enquote{language of thought}.
	Within this language of thought, symbols remain detached from specific sensory modalities: the concept of a bird can be evoked through vision, hearing, or touch, but it may also be activated amodally through language, or even through abstract reasoning.
	Secondly, the symbols can be arbitrarily bound together, even in spite of strong evidence to the contrary: symbols can be bound if there are opposing semantic co-occurence statistics, e.g., ``a sulfur-smelling bird'', or even logical/semantic contradiction, e.g. ``a square circle''.
	Symbols are discrete in the sense that the symbols themselves are not graded or continuous, but rather exist as distinct entities in memory that can be combined and manipulated in a rule-governed manner, explaining why so many of our concepts and thoughts are categorical and devoid of graded nuance.
	These properties of symbols allow for a high degree of combinatorial freedom, as symbols can be combined in a flexible and arbitrary manner, allowing for the generation and comprehension of an unbounded number of complex ideas in a computationally effecient manner \citep{frankland_language_2019,marcus_algebraic_2001,chomsky_language_2006,sommers_bundle_nodate}.
	Another empirical observation that contrasts with PACED is that rather than reading books or watching movies in a parallel fashion -- imagine collapsing an entire story into a single input state, reading it all at once -- we instead read and watch events unfold one word or frame at a time, incrementally building up a mental model of the story as we go along \citep{sanford_memory-based_2005,garrod_incrementality_1995,lewis_computational_2006}.
	This is not merely an artifact of time-bound sensory input: even when we process a complicated still image containing many interacting objects or parts, we do so serially, moving our eyes around the image to build up a mental model of the scene \citep{henderson_human_2003,rayner_eye_2009}.
	In other words, an important part of cognition seems to be Serial, (semantically) Arbitrary, and Discrete (SAD), and symbolic systems have traditionally been considered as good models of these capacities.

	In line with this reasoning, a large body of literature has argued that there exist some phenomena that are very hard to explain with only a connectionist system \citep{lake_generalization_2018,loula_rearranging_2018,van_der_velde_neural_2006,marcus_algebraic_2001}.
	Moreover, some of these phenomena are easy to explain with symbolic systems: reasoning using predicate-based logic, especially when using language, is a prime example of this.
	Furthermore, variable assignment -- a key feature of symbolic systems -- trivially explains one-shot learning, i.e. the ability to learn a new concept after observing only a single example.
	Connectionist systems often require many repetitions to do so \citep{hadley_problem_2009,marcus_algebraic_2001,lake_one-shot_2014}, presumably because they need to adjust many parameters (i.e. the curse of dimensionality).
	Connectionist systems also struggle with the problem of two (also known as the problem of interference or as pattern separation), where two distinct concepts with similar overlapping features are confused with each other \citep{van_der_velde_neural_2006,oreilly_complementary_2014,fodor_connectionism_1988,jackendoff_foundations_2003} — though see \citet{sommers_bundle_nodate} for nuance.
	In symbolic systems, variable assignment easily solves the problem of two, since a symbolic system can simply store concepts into distinct, separate variables, no matter how similar they are.
	Finally, compositionality is an inherent feature of many symbolic systems, allowing for the flexible generation and comprehension of an unbounded number of complex ideas \citep{frankland_language_2019,lake_generalization_2018}.
	In fact, it is not uncommon for connectionist systems to be criticized for their lack of compositional freedom, as they struggle to allow the same degree of combinatorial freedom that symbolic systems do \citep{van_der_velde_neural_2006,fodor_connectionism_1988,lake_generalization_2018}.
	More broadly then, it seems to be the case that symbolic systems have certain advantages over traditional connectionist systems -- even if the difference between connectionist and symbolic systems is not as clear-cut as assumed here (see \citet{sommers_bundle_nodate} for a discussion).
	However, both types of systems are Turing complete \citep{turing_computable_1937,church_unsolvable_1936,neumann_first_1993,hornik_multilayer_1989,siegelmann_computational_1992}, meaning that every phenomenon/problem that can be solved by either a connectionist or a symbolic system, given enough time and resources, can also be solved by the other.
	The difference, then, does not lie in the computational capabilities provided by these systems, but rather in the constraints and assumptions imposed by them, and the algorithmic approaches that this allows for.
	Symbolic systems, rather than operating over a high-dimensional continuous space where different values are computed in parallel, can instead operate over a compressed discrete space where only the relevant items are stored and updated, in a serial fashion.
	As argued above, this compression makes that a purely symbolic system is not well-suited to capture the PACED nature of cognition.
	However, it may be exactly this lossy compression that allows symbolic systems to tractably solve certain phenomena/problems that are inefficient or even intractable for a purely connectionist system, such as one-shot learning, pattern separation, and compositionality.
	As such, a combination of both systems may be needed to capture both the PACED and SAD aspects of cognition.
	For instance, one may imagine a PACED low-level system capable of representing fine-grained, high-dimensional semantic information, coupled with a SAD high-level system that can operate over the compressed representations of the low-level system.

	All models of cognition explain a particular subset of phenomena -- the phenomena they have targeted -- while often struggling with others. 
	Oftentimes, the phenomena that a model struggles with are partly inherited from which tradition it is from, e.g., symbolism or connectionism.
	We pose that any satisfactory model of cognition should be able to explain phenomena that are explained by the superset of both traditions, since a model that can explain a wide range of phenomena is less likely to be overoptimized for a particular subset of tasks at the expense of others.
	The PACED and SAD aspects of cognition that we introduced here are still high-level descriptions, and testing them requires further specification.
	In order to address this, we propose a set of phenomena that we call Symbol Recombination, Retention, and Resolution (S3R).
	This set of phenomena is not exhaustive, nor is it static: our aim has been to keep collecting phenomena that are hard to model by any (but not necessarily all) existing models, so as to ensure that our model is as comprehensive as possible.
	This set of phenomena includes semantics, compositionality, retention, resolution, and control.
	Each of these categories detail what we believe to be important properties of cognition in some domain, and the phenomena that arise when these properties are not present in a model.

	The brain stores and processes a vast network of semantic information.
	One key property of this semantic network is that it is context-dependent (Phenomenon 1): the meaning of a concept is embedded in a semantic graph whose activation patterns are determined by the context in which the concept is used \citep{barsalou_situating_2008}.
	For instance, priming studies reveal that the activation of a concept can be influenced by the activation of related concepts, even when these concepts are not explicitly mentioned \citep{schacter_priming_1998}.
	Without such context-dependency, the evoked meaning of concepts would always be the same and would retrieve too much unnecessary information: rather absurdly, one might imagine an avid bird-watcher recalling all of their knowledge about the appearance and behaviour of ducks upon being presented with a plate of roasted duck.
	Graded inference (Phenomenon 2) is another property that is crucial for a semantic network: the ability to reason about the world in a way that is not categorical, but continuous \citep{rosch_natural_1973,taylor_prototype_2011}.
	This starts with the concepts themselves, whose information is stored in a distributed manner across the network, and whose connections are not binary \citep{rosch_natural_1973}.
	Priming studies further support this idea, as they show that the activation of a concept can be influenced in a subtle and graded manner \citep{schacter_priming_1998}.
	Though one may argue a certain level of black-and-white thinking is present in humans, it is evidently not the case that all of our reasoning is categorical, and a model that is capable of exhibiting graded and context-dependent reasoning is more likely to be able to capture the complexity of human cognition.

	One large claim in cognition and, in particular, language, is that of compositionality: the idea that complex ideas can be built from simpler ones in a systematic and rule-governed manner \citep{fodor_connectionism_1988,frankland_language_2019}.
	We will distinguish between two types of compositionality: flat compositionality (Phenomenon 3), where concepts are combined in a single step, and recursive compositionality (Phenomenon 4), where concepts are combined in a hierarchical manner.
	Flat compositionality is tightly connected to the binding problem in connectionist literature or (dynamic) variable binding in symbolic literature, and it is an ongoing question in neuroscience and AI \citep{feldman_neural_2013,hummel_getting_2011,greff_binding_2020}.
	\citet{feldman_neural_2013} clearly relates the two, stating that dynamic variable binding is required to solve the (neural) binding problem (Phenomenon 5), which he further distinguishes into four (related but) distinct subproblems, including visual feature-binding, and dynamic variable binding.
	A solution to the binding problem is required to allow arbitrary combinations of concepts and preventing interference between similar items (Phenomenon 6).
	Without a solution to the binding problem, we would require brute force enumeration of all possible combinations, which is intractable due to the combinatorial explosion of options \citep{becke_neural_2015, van_der_velde_necessity_2015}.
	When considering all possible combinations for N concepts, there are $2^N$ possible bindings.
	For example, for only 25 concepts, there are already more than 33 million possible combinations.
	Given that humans easily have more concepts than that, this approach would quickly exceed the number of neurons in the brain (or, for that matter, the number of atoms in the universe).
	In particular, this has been a hard problem for connectionist systems, which often struggle with the binding problem and the tractable implementation of compositionality \citep{van_der_velde_neural_2006,fodor_connectionism_1988}.
	
	However, once we have a tractable mechanism for binding, we can effectively repeat the same process to combine the results of previous combinations, leading to recursive compositionality.
	This implies that the selected solution to the binding problem must be able to handle multiple levels of binding.
	Recursive compositionality is a key feature of many symbolic systems, notably language, and it allows for the \enquote{flexible generation and comprehension of an unbounded number of complex ideas} \citep{frankland_language_2019,chomsky_language_2006}.
	One example of this, often used in language, is role-filler independence (Phenomenon 7): it states that roles (e.g., one-place predicates such as \enquote{x walks}, \enquote{x loves (active)}, \enquote{x is loved (passive)}) and fillers (e.g., \enquote{John}, \enquote{Mary}, \enquote{the dog}) should be computationally kept separate such that they can be combined in a flexible and arbitrary manner \citep{hummel_solution_2004,hummel_getting_2011}.
	Without recursive compositionality, we would be limited to a single level of combination, e.g., being able to express that \enquote{John loves Mary}, but not that \enquote{John loves the dog that Mary loves}.
	Thus, in order to represent the complex capabilities of human cognition, as seen in language and other domains, we believe that a model should be able to account for both flat and recursive compositionality.

	The ability to retain information over time is crucial for any cognitive system, as it is a prerequisite for the accumulation of knowledge and the building of complex ideas.
	Memory itself has been implemented in many ways in cognitive models, and so the ability to retain information itself is not generally a hard phenomenon to explain.
	However, the way in which information is acquired and retained is harder to model for some model classes.
	Connectionist systems for example, with their many parameters, often require many repetitions to learn a generalizable solution \citep{fei-fei_one-shot_2006,burgess_one-shot_2017,vinyals_matching_2016}.
	In contrast, humans show clear evidence of one-shot learning (Phenomenon 8), also known as fast mapping in the case of word learning \citep{carey_acquiring_1978,bion_fast_2013,spiegel_rapid_2011} — if we would not, we would never be able to learn someone's name after only a single introduction.
	Another example is discourse incrementation (Phenomenon 9), where the understanding of a discourse is built up incrementally, requiring both the retention and integration of information over time \citep{sanford_memory-based_2005,garrod_incrementality_1995,lewis_computational_2006}.
	If we were unable to perform discourse incrementation, we would not be able to create a mental image of a character described in a novel, or to gradually build up understanding about a complex concept.
	Working memory is a key component of this process (Phenomenon 10), as it allows for the temporary storage and manipulation of information over short periods of time \citep{baddeley_working_2010}.
	Presumably, if all information were to be stored in memory, we would quickly accumulate so much information that we would no longer be able to see the forest for the trees.
	In that vein, forgetting (Phenomenon 11) is a crucial part of working memory \citep{lewis_computational_2006}: if we were unable to forget information, our working memory would quickly become overloaded, and we would be unable to focus on the most relevant information.
	As such, to be able to learn quickly and flexibly, and to be able to build up complex ideas over time, we should expect both gradual and one-shot learning, both short- and long-term memory, and both retention and forgetting to be present in any cognitive model.

	Of course, retention is only half of the story: the ability to retrieve information is just as important.
	We know that the brain can retrieve memories using content-addressable memory (CAM, Phenomenon 12), where the activation of a partial memory can lead to the retrieval of the full memory \citep{lewis_computational_2006,mcelree_memory_2003,parker_cue-based_2017}, effectively completing the stored pattern.
	This idea is very intuitive from a connectionist lens, as interconnected units can activate each other in a distributed manner.
	However, its implementation in more symbolic paradigms is less clear.
	Without the ability to retrieve information in this way, one would need to enumerate all memory traces and compare them to the partial memory, which would quickly become computationally costly as the number of memories grows.

	Additionally, it is unclear what happens in problem of two cases (Phenomenon 13), where two similar but distinct memory items can cause interference \citep{van_der_velde_neural_2006,jackendoff_foundations_2003}.
	A classic example is the sentence ``the little star is to the left of the big star''.
	Such sentences have historically been considered hard problems for connectionist systems, as they struggle to distinguish between similar items \citep{fodor_connectionism_1988,van_der_velde_neural_2006}.
	One way in which connectionist systems can solve this problem is through hippocampal pattern separation (Phenomenon 14), as argued by complementary learning system theory \citep{mcclelland_why_1995,mcclelland_considerations_1996,oreilly_complementary_2014,kumaran_what_2016}.
	By using a combination of expansion recoding, strong thresholding, lateral inhibition, and coincidence detection of memory, the hippocampus can effectively separate similar but distinct patterns \citep{marr_theory_1970,albus_theory_1971,cayco-gajic_re-evaluating_2019}.
	This transforms a semantically continuous input space into an output space where semantic differences are maximally separated, leading to (more) discrete representations.
	The prevention of interference between memory items also requires serializing the memory traces through cognitive control \citep{musslick_rationalizing_2021}, so that only one memory can be activated at a single time (Phenomenon 15).
	Besides activating singular memory traces, the brain can also store and retrieve sequences of memories (Phenomenon 16), allowing for the retention of temporal information and the replay of past events composed of multiple time slices \citep{pavlides_influences_1989,buhry_reactivation_2011,buzsaki_space_2018}.
	Without the presence of these retrieval mechanisms, our models would not be able to match the rich and complex capabilities of human memory.

	Somewhat counterintuitively, an important aspect of memory retrieval is the ability to suppress this retrieval (Phenomenon 17), often vital to prevent interference between some memory and other cognitive processes or memories \citep{levy_inhibitory_2002,herrmann_control_2001}.
	This is especially important when forming a new memory that partially overlaps with an existing one but needs to be stored separately \citep{musslick_rationalizing_2021,van_kesteren_how_2012}.
	Without a mechanism to suppress retrieval, the new information would simply be merged with the existing memory, leading to a loss of information and a potential confusion between the two items.
	Suppression can also help explain phenomena like that of correlation violation (Phenomenon 18), where learnt associations that are stored in long-term memory are violated by new information, but especially connectionist systems are unable to suppress the retrieval of these associations to allow for the learning of new ones \citep{van_der_velde_neural_2006,puebla_relational_2021}.
	All of these phenomena relate to the ability to control memory retrieval, and show that simple retrieval mechanisms are not sufficient to capture the complexity of human memory.

	However, the ability to control the memory system extends beyond just retrieving memories.
	Different task contexts require different operations to be performed, and the memory system should not just be able to perform each of these operations, but also to switch between them in a controlled and selective manner. 
	For instance, when a friend tells you that the person speaking to the bartender is a friend of theirs, you need to be able to identify the bartender, see who is speaking to them, and then assign to this person the role of being a friend of your friend.
	In this example, clear control is required over and above passive memory retrieval of the appropriate word meanings: the system needs to actively prevent assigning the role to the bartender.
	This role assignment instead involves controlled juggling of the different individuals and encoding/manipulating the memory traces of the individuals and their attributes (Phenomenon 19).
	This ability to manipulate memories distinguishes working memory from short-term memory, as working memory is not just a passive store of information, but an active system that can be used to perform computations on the stored information \citep{baddeley_working_2010}.
	In order to perform manipulations, the memory system should be able to distinguish between novel entities and existing ones (Phenomenon 20) \citep{bion_fast_2013}, and in order to solve the problem two, it should also be able to detect when a memory is ambiguous (Phenomenon 21).
	Detection of these different situations can then be used to selectively suppress parts of the memory system, so that only the relevant information is retrieved or manipulated, and that it is stored in the right place.
	Another operation that requires control is the retrieval of arbitrary memories, i.e., not through the content but by virtue of holding a particular position in the memory system.
	This is called addressable read-write memory (Phenomenon 22), and it is a key feature of working memory that allows for the selective manipulation of memories \citep{gallistel_memory_2009,atkinson_human_1968,awh_working_2025}.
	These operations are agnostic to the content of the memories, and are instead focused on the management of the location of memories and their arbitrary retrieval.
	Altogether, without control, the complex memory mechanisms outlined in the previous sections would be unable to function properly, making it especially complicated to construct a model that can explain all of these phenomena — yet vital in order to capture the full complexity of human cognition.

	There are existing models that combine the advantages of both connectionist and symbolic systems, and that can explain a wide range of cognitive phenomena. 
	One such model is the Adaptive Control of Thought-Rational (ACT-R) model, which combines a connectionist memory system with a symbolic control system \citep{anderson_how_2009}.
	However, these models are often criticized as duct-tape AI, as they are often not fully embedded in a connectionist system, nor neurobiologically plausible \citep{eliasmith_how_2013}. 
	Other hybrid models like vector symbolic architectures (VSAs) and the Semantic Pointer Architecture (SPA) \citep{stewart_spaun_2012} use either tensor products \citep{smolensky_tensor_1990}, circular convolution \citep{plate_holographic_1995} or matrix multiplication \citep{gallant_representing_2013} to combine symbolic and connectionist representations. 
	These models provide promising ways to explain S3R phenomena \citep{gayler_vector_2004}, but their binding operations are mathematical abstractions without clear biological implementations.

	Although influenced by these approaches, we propose a more principled hybrid approach to solving the S3R phenomena, in which all components are fully embedded in the same system and implemented in a neurobiologically plausible manner.
	We call this proposal the Symbolic SubSystem Hypothesis.
	It makes the following three claims:
	\begin{enumerate}
		\item The brain is a Parallel and Distributed processor, constisting of neurons whose Associative synaptic connections and Continuous activation patterns are used to represent and process Embodied semantic information, i.e. the brain can perform PACED computations. 
		\item The brain at the same time also has a symbolic subsystem implemented in its neural wetware that extracts Discrete attributes from this continuous semantic space, allowing the brain to Serially perform symbolic operations such as (relational) composition of Arbitrary attributes, i.e., the brain can perform SAD computations.
		\item This symbolic subsystem is an integral part of and works in tandem with the other parts of the brain, thereby combining the advantages of both PACED and SAD computation.
	\end{enumerate}
	Consider the following analogy: a computer is a symbolic system capable of performing arithmetic operations on numbers by performing read/write operations onto discrete memory registers.
	However, computers are often used to simulate connectionist systems, e.g., (spiking) neural networks, in a way that is clearly considered accurate enough for the fields of computational neuroscience and AI\@.
	Implementational connectionism proposes the reverse: that the brain is fundamentally a connectionist system, but that this connectionist system implements the operations required to perform symbolic computations \citep{pinker_language_1988}.
	The Symbolic SubSystem Hypothesis is highly similar to implementational connectionism, but more specific: it does not just propose that the brain is a neural/connectionist system implementing a symbolic system, but that this symbolic system is a subsystem of the brain, which works in tandem with the larger neural/connectionist system.
	The algorithm of the brain is thus not purely symbolic, but is instead a symbolic-connectionist hybrid, fully implemented in neural/connectionist hardware.

	To be clear, the Symbolic SubSystem Hypothesis is deliberately vague and does not make any claims about many of the specifics, such as whether the symbolic subsystem is distributed across the brain or localized in a particular area, or whether it is used for all cognitive tasks or only for some.
	It merely posits that current evidence suggests that the brain implements such a system in a connectionist manner, that such an implementation would allow for simple explanations of many of the S3R phenomena, and that the brain performs symbolic computations in some form.
	The model we will propose is a specific instantiation of this hypothesis, and while capable of interesting symbolic computations, does not yet provide the full story.
	Thus, the Symbolic SubSystem Hypothesis should be taken as somewhat distinct from the model we will propose going forward, and as a more general claim about the nature of cognition.

	Prior to this work, we have developed a functioning cognitive system for reference comprehension that explains a large part of the S3R phenomena \citep{sommers_bundle_nodate}.
	It explains the task of anaphoric reference (coreference) in linguistics, where a model is required to understand a sentence (in an artificial language) and to use this understanding to resolve what entities are being referred to.
	However, this system does not yet provide a detailed neurobiological implementation: there are several separate components, only one of which is connectionist in nature, and the control system is fully hard-coded through conditional logic.
	Although we do provide potential venues for the neurobiological implementation of the different components, this model still falls prey to many of the criticisms levelled at hybrid models, such as the homunculus problem and the duct-tape AI criticism.
	As such, we aim to provide a more specific, fully integrated neural model that can explain the S3R phenomena, and that can be used as a reference for future models of cognition.

	Another model that has similar goals is the conjunction neuron model by \citet{manohar_neural_2019}, which implements a rate-coding model of symbol-like working memory slots or registers.
	This model uses a pool of neurons, called conjunctive neurons, that have facilitating synapses with all attributes, meaning they can bind to them.
	By using lateral inhibition, the model can perform a winner-take-all operation, effectively selecting a single memory trace from a pool of candidates.
	However, this model struggles to explain several S3R phenomena: it cannot reliably resolve cases with multiple candidates that have partially overlapping attributes (the problem of two), it cannot gradually build up memory traces over time (discourse incrementation), and there are several other semantic tasks that it cannot perform, such as transitive inference.
	We hypothesize that these limitations stem primarily from the lack of a control system that can distinguish between novel and existing memories, dynamically allocate new memory slots while protecting existing ones, and suppress the whole system during other activity.
	Despite this limitation, this model is a great example of a fully connectionist and neurobiologically plausible model capable of solving a broad set of S3R phenomena.
	Moreover, it explains a wide range of findings in the working memory literature and thus provides a good starting point.

	We fine-tune and extend the conjunction neuron model with a control system to explain a large subset of S3R phenomena.
	This model, which we call the Rate-Coded Bundle Memory (RCBM), is a rate-coding model similar to that used by \citet{manohar_neural_2019} in their conjunctive neuron model.
	However, it contains several additional mechanisms that allow it to explain a broader set of S3R phenomena, the most important of which are required to implement a control system.
	In reference to our earlier work \citep{sommers_bundle_nodate}, RCBM functionally implements Bundle Memory (BM), which explains many S3R phenomena, but in contrast to our earlier work it is fully embedded in a connectionist system.
	Importantly, we argue that the combination of mechanisms available to RCBM enables the formation of a Symbolic Subsystem, and that it is exactly the presence of this subsystem that allows RCBM to explain such an extensive range of S3R phenomena.

	Another driving force for this work is our novel approach of operationalizing top-down control in a rate-coding model.
	In the cognitive neuroscience literature, top-down control (of memory) is widely recognized as being an important aspect of (higher) cognition.
	This interplay between memory and the control of memory is a central theme in the neurobiology of language as well: the Memory, Unification, and Control (MUC) model, for example, explicitly distinguishes the linguistic knowledge stored in memory from the control processes that regulate how it is accessed and combined \citep{hagoort_muc_2013}.
	Such control processes are used to guide attention, to suppress irrelevant information, and to select the most relevant information for the task at hand.
	Yet, the use of the word ``control'' is often used in a metaphorical sense, and it is not clear how this control is implemented in the brain, which invites the criticism that cognitive neuroscientists commit a homunculus fallacy, i.e., by delegating all hard problems to the homunculus supposed to control everything the brain does, they are just pushing the problem of cognition to a different level \citep{monsell_control_2000}.
	By providing a rate-coding model of top-down memory control, we aim to show that a dedicated top-down control system should not be seen as a distinct entity but as an integral part of the brain, and we show how one can implement such a system through specific neural circuitry.

	\section{Methods}\label{sec:methods}

	\begin{figure*}
		\centering
		\includegraphics[width=\textwidth]{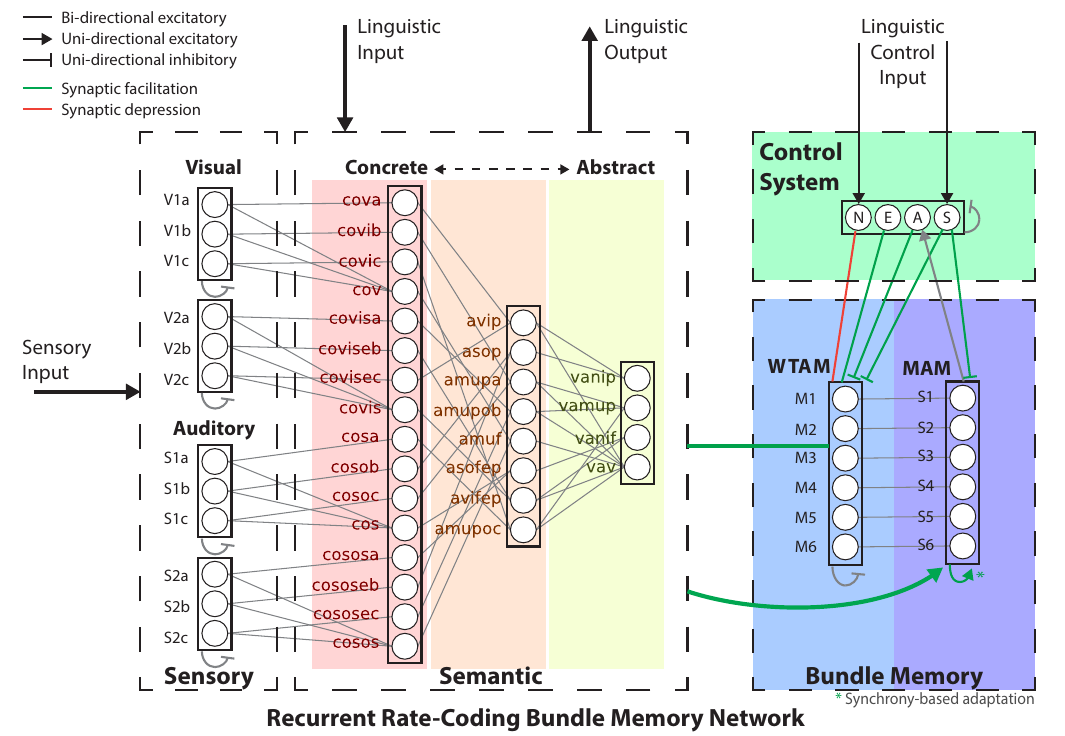}
		\caption{
			\small
			The RCBM model consists of three main components: a semantic system, a control system, and a memory system.
			The semantic system (left and center) consists of both visual and auditory sensory feature neurons (left) and conceptual feature neurons (center).
			Each neuron in the semantic system can directly receive external input through corresponding sensory or linguistic input.
			Linguistic input consists of multiple sentences in an artificial language, presented word-for-word.
			Each word is represented by a conceptual feature neuron, and these conceptual feature neurons are organized in increasingly abstract layers, with the rightmost layer representing the most abstract concepts.
			Each non-sensory semantic neuron can directly output corresponding artificial words for the model to produce a response.
			The control system (top right) consists of four control neurons: a novelty detection neuron (N), an existing memory detection neuron (E), an ambiguity detection neuron (A), and a memory suppression neuron (S).
			Special linguistic inputs can directly activate the novelty detection neuron (determiner, e.g., \textbf{a} man v.s.\ \textbf{the} man) and the suppression neuron (end-of-sentence token).
			The memory system (bottom right) consists of two memory pools: a Winner-Take-All Memory pool (WTAM) and a Multiple Activation Memory pool (MAM).
			Each memory pool consists of a set of memory neurons, which are connected to the conceptual feature neurons and have pairwise connections between them.
			The control system interacts heavily with the memory system, as it is responsible for activating the appropriate memory neurons and suppressing the memory system when necessary.
		}\label{fig:rcbm}
	\end{figure*}

	In prior work \citep{sommers_bundle_nodate}, we developed a model of working memory called Bundle Memory (BM), named after the thesis by David Hume that objects are nothing but bundles of ideas (attributes) \citep{hume_treatise_2000}.
	This model was designed to explain the S3R phenomena, and consists of three main components: a control system, a bundle memory module, and a semantic system.
	Here, we implement this model using a rate-coding model, to embed all components in a single connectionist system.
	In order to achieve this, we provide neurobiologically plausible implementations of the control system and bundle memory modules.
	We call this model the Rate-Coded Bundle Memory (RCBM) model.

	According to the framework constructed in our previous work \citep{sommers_bundle_nodate}, all three components of the model are crucial for the functioning of a system capable of explaining the S3R phenomena.
	The semantic system is responsible for the representation of concepts and their relations, and it allows for graded inference, context-dependence, and groundedness.
	The BM system is responsible for the retention and manipulation of information, and it allows for one-shot learning and dynamic variable binding.
	It is implemented through monosynaptic binding, which allows for the creation of short- to medium-term memory traces.
	Finally, the control system is responsible for the dynamic allocation of memory slots, the distinction between novel and existing memories, and the suppression of the memory system during other activity.
	In RCBM, we implement all parts within the same rate-coding model, with specialized circuits and connection types to distinguish their functioning (see \autoref{fig:rcbm}).

	In the RCBM model, we provide a more detailed implementation of all three systems.
	Here, the semantic system consists of a hierarchically organized set of attribute neurons, with connections to visual and auditory sensory neurons (\autoref{fig:rcbm}, left).
	Each of these attributes can be externally activated through either sensory input (left) or linguistic input (top).
	The BM system consists of two memory pools: a Winner-Take-All Memory (WTAM) pool and a Multiple Activation Memory (MAM) pool (\autoref{fig:rcbm}, bottom right).
	These pools have pair-wise connections between them, and each pair of memory neurons constitutes a single bundle (memory trace).
	Each attribute neuron in the semantic system is connected to all memory neurons in the WTAM and MAM pools through facilitating synapses that allow for monosynaptic binding.
	In this way, the model can store attributes in specific bundles by binding to their corresponding memory neurons.
	The control system consists of four control neurons.
	These neurons are responsible for identifying whether the current input is novel (N) or belongs to an existing memory (E), whether there are multiple candidate memories (A), or whether the memory system should be suppressed/cleared (S) (\autoref{fig:rcbm}, top right).
	These control neurons interact heavily with the memory system to allocate new memory bundles, to protect existing ones, and activate the appropriate bundle upon retrieval.
	
\begin{figure*}
		\centering
		\includegraphics[width=\textwidth]{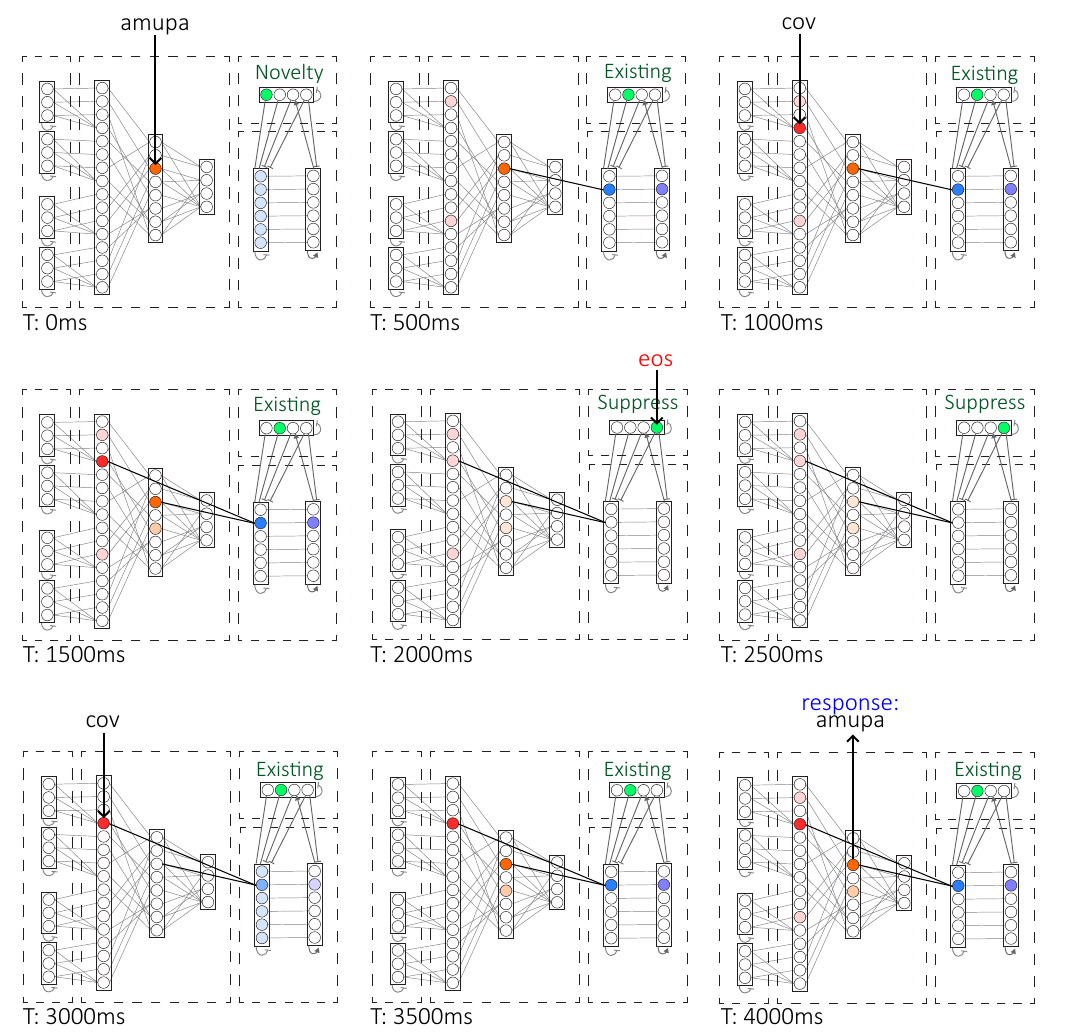}
		\caption{
			\small
			The RCBM model stores and retrieves information to and from a bundle in response to an input sequence.
			The artificial language sequence (\enquote{Amupa cov. Cov ...?}) may be compared to a sentence like \enquote{There is a black cat. The cat is ...?}, where the ellipsis indicates that the model is expected to retrieve the information that was stored in the first sentence.
			Specifically, in response to the first word (amupa, see \autoref{fig:rcbm}), the corresponding attribute neuron is activated, and the control system detects that this is a novel memory ($T = 0\text{ms} - 500\text{ms}$).
			The control system allocates a new memory slot in the Winner-Take-All Memory (WTAM) pool, which then activates the existing control neuron.
			The semantic information is then bound to the WTAM and MAM neurons belonging to the same bundle ($T = 500\text{ms} - 1000\text{ms}$).
			After the first word, the model receives a second word (cov), again activating the corresponding attribute neuron which is then also bound to the WTAM and MAM neurons of the current bundle ($T = 1000\text{ms} - 2000\text{ms}$).
			The model then receives an end-of-sentence signal (EOS), which causes the control system to suppress the memory system and to prepare for the next input sequence ($T = 2000\text{ms} - 3000\text{ms}$).
			Finally, the model receives the second word again (cov), which causes the control system to detect that this is an existing memory and to reactivate the corresponding bundle ($T = 3000\text{ms} - 3500\text{ms}$).
			This bundle will then lead to the reactivation of the other semantic information that was bound to it (amupa), and the model will be able to retrieve the information that was stored in the bundle ($T = 3500\text{ms} - 5000\text{ms}$).
		}\label{fig:rcbm_sequence}
	\end{figure*}

	Together, these components allow the RCBM model to retain bundles of information over time, to retrieve them when needed, and to manipulate them in a controlled manner.
	Upon being presented with sensory or linguistic input, the corresponding semantic attribute neurons are activated, and the control system detects whether this is a novel memory or an existing one (see \autoref{fig:rcbm_sequence} for example sequence).
	If the input is novel, the control system allocates a new memory slot in the WTAM pool.
	If there already exists a bundle that matches the input, that memory emerges from the WTAM pool through winner-take-all dynamics.
	In cases where there are multiple candidate memories, the control system detects this through the MAM pool and awaits disambiguating information before proceeding.
	Semantic information that is bound to the currently active bundle is reactivated through connections back to the semantic system from the WTAM pool.
	Any new semantic information is simultaneously bound to the respective WTAM and MAM neurons.
	In between sequences of words, to prevent interference between bundles, the memory system can be cleared through the suppression neuron.
	The interactions between these systems is what allows the RCBM model to turn sequences of sensory and linguistic input into bundles of information in a controlled and organized manner.

	\subsection{Rate Coding Model}

	Our motivation for using a rate coding model is that it balances computability with neurobiological plausibility.
	Rate coding models are directly informed by observations from neural recordings, and are based on the idea that the firing rate of a neuron can be used to encode information.
	At the same time, they significantly reduce computational requirements compared to spiking neural networks, which makes them more suitable for large-scale simulations.
	Our RCBM simulation is split into two main parts per time step: signal propagation and neural activation.

	During signal propagation, all neurons propagate their signals to all neurons they are connected to, implemented in the form of a matrix multiplication.

	We use leaky neurons, which means that the new state of a neuron is determined by the old state ($\vec{r_t}$), and some exponential decay through which the signal returns to baseline over time:

	\[\vec{r}_{t+1} = f(\vec{r}_{change} + \vec{r_t} + b) - b\]

	Where $\vec{r}$ is a vector containing the firing rates of all neurons, $f$ is a neural activation function, $b$ is the baseline activation of the neurons.
	The activation function we use is a linear function clipped between 0 and 1, with 0 representing no firing and 1 representing the maximum firing rate.
	In this equation, $\vec{r}_{change}$ is the change in firing rate before applying the activation function and baseline correction, calculated as:
	\[\vec{r}_{change} = \gamma W \vec{r_t} - d_{r}\vec{r_t} + \vec{\zeta}_{t}\]

	Here, the first part ($\gamma W \vec{r_t}$) represents the incoming signal from other neurons, where $\gamma$ is a scaling factor that determines the general signal propagation strength and $W$ is the weight matrix which defines the connections between neurons. 
	Excitatory and inhibitory connections are determined by the sign of the scaling factor.
	The second part ($- d_{r}\vec{r_t}$) represents the decay of the firing rate, where $d_{r}$ is the rate at which the firing rate decays to baseline.
	The third part ($\vec{\zeta}_{t}$) represents some noise in the system, sampled from a zero-mean normal distribution with a standard deviation equal to noise parameter $\zeta$.
	For most neurons, assuming that the baseline firing rate accounts for the overall activity, we set the noise to 0.
	Only for processes that rely on stochasticity, we explicitly add noise to the system.

	\subsection{Conjunction Neurons as the Basis of Bundle Memory}

	Beyond the general rate coding model, which dictates the way in which neurons interact with each other, we also implement several specialized circuits and connection types to distinguish the functioning of the different components of the model.
	We use the conjunction neuron model by \citet{manohar_neural_2019} as a basis for our model, as it provides a rate coding model of working memory that can already explain a large number of S3R phenomena.
	Primarily, we use the idea of conjunction neurons, which have facilitating synapses with all attributes allowing them to bind to them, to form a Bundle Memory-like model of working memory.
	Their model also incorporates basic mechanisms for memory allocation and selection, which we extend with a more sophisticated control system to explain a broader set of S3R phenomena.

	The primary ingredient to enable the formation of dynamic memory traces in our model is Fast Hebbian Plasticity (FHP), which has been proposed for neurobiological computational models of working memory \citep{fiebig_spiking_2017,fiebig_indexing_2020,sandberg_working_2003,durstewitz_neurocomputational_2000,manohar_neural_2019}.
	Hebbian learning is a mechanism by which the connection strength between two neurons increases when they fire at the same time \citep{kempter_hebbian_1999,hebb_organization_2005}.
	We implement FHP in the memory and control modules of our model, allowing for dynamic changes in the connectivity between neurons:

	\[\Delta W = \eta \mathcal{H}(\vec{r}_{\text{source}}, \vec{r}_{\text{target}}) \]

	Where $\eta$ is the FHP rate and:

	\[\mathcal{H}(\vec{r}_{\text{source}}, \vec{r}_{\text{target}}) = g(\vec{r}_{\text{source}}) \otimes g(\vec{r}_{\text{target}})\]

	Where $g$ is some FHP function, $\vec{r}_{\text{source}}$ and $\vec{r}_{\text{target}}$ are the firing rates of the source and target neurons, and $\otimes$ is the outer product.
	We disallow self-connections, as these would lead to neurons always reinforcing their own activation.
	By default, and in the original model by \citet{manohar_neural_2019}, the FHP function is the identity function, meaning that the connection strength increases linearly with the activation of the neurons.
	However, we also use different FHP functions in our model for specialized neurons, to allow for more complex learning dynamics.
	Such specialized neurons will be discussed in more detail in later sections.
	By implementing Hebbian learning, we create the conditions for a system that can dynamically adapt its connectivity to represent new information.

	In \citet{manohar_neural_2019}, learning of new memory traces occurs through a pool of neurons called conjunctive neurons, each of which has facilitating synapses to all attribute neurons, allowing them to dynamically bind to them.
	These neurons very much resemble a certain type of neurons in our model, and as such we use a very similar mechanism to implement these here.

	We use a strong FHP rate ($\eta$) for connections between the memory neurons to the attribute neurons, meaning that the connection strength increases significantly when both neurons are fully active, allowing these neurons to effectively bind to the attributes they are active with.
	These connections are bidirectional, allowing for both reactivation of attributes from memory, and for retrieval of memory from attributes.
	However, the connections from memory neurons to attribute neurons are stronger than the other way around, meaning that memory retrieval occurs at a slower rate than the reactivation of attributes that are part of an active memory trace.

	To enable competition between memory neurons, like \citet{manohar_neural_2019}, we implement lateral inhibition between the memory neurons, which leads to a winner-take-all paradigm.
	Because of this, we will hereafter refer to the primary memory neurons as Winner-Take-All Memory (WTAM) neurons (\autoref{fig:rcbm}, bottom right).
	This lateral inhibition is implemented as a negative scaling factor ($\gamma$) between the memory neurons, which ensures that when one memory neuron is activated, the other memory neurons are suppressed.
	As such, when multiple memory neurons are activated, the lateral inhibition ensures that, in the end, only one of them is selected.
	This allows for the selection of a single memory neuron when attributes are activated that are associated with multiple memory neurons, where the memory neuron most similar to the activated attributes is selected.
	Taken together, this is what allows for Content Addressable Memory (CAM) in our model, where the activation of a partial memory can lead to the retrieval of the full memory.

	To allocate a new memory neuron when no other distinguishing attributes are present, i.e., when a novel memory trace is being formed, we require the noise parameter $\zeta$ to introduce random noise to the system \citep{manohar_neural_2019}.
	This noise allows for the selection of a random WTAM neuron when no other distinguishing attributes are present, effectively allowing for the assignment of random unassigned WTAM neurons to novel bundle memories.

	\subsection{Increasing Network Stability}

	The original model by \citet{manohar_neural_2019} uses a relatively simple rate-coding model to implement their conjunction neurons.
	This model works well for the tasks they set out to explain, but it struggles with some of the more complex S3R phenomena, such as discourse incrementation and the problem of two.
	To explain these phenomena, we need to extend the model with more specialized neural mechanisms that allow for the assignment and retention of information more reliably, and that allow the model to act in a more discrete and symbolic fashion.
	We will discuss these specialized neural mechanisms in more detail in the following sections.

	First of all, we implement decay to our Hebbian weights to enable gradual forgetting of memory traces, regardless of network activity.
	This is important to prevent the network from becoming saturated with memory traces, and to allow for the formation of new memory traces over time.
	The original model by \citet{manohar_neural_2019} relied on active unlearning over time, where activations below baseline would lead to negative Hebbian learning and thus to the reduction of the connection strength.
	We implement a more passive form of decay, where the connection strength of Hebbian weights decreases over time to some specified baseline:

	\[\Delta W = (W - w_0) d_w\]

	Where $w_0$ is the baseline connection strength, and $d_w$ is the weight decay rate.

	For our WTAM neurons, we used non-zero baseline for connections from attributes to memory neurons.
	Without this baseline, unbound attributes would not be able to activate any memory neurons, and as such the model would not be able to form new memory traces.
	Connections back from memory to attribute neurons, however, do not require a baseline, and were hence set to 0.

	To further increase the robustness of our Bundle Memory model, we extend our Hebbian learning with sigmoidal synaptic activation thresholds.
	This means that binding only takes place under strong pre- and post-synaptic activation, which is desirable to support the binary nature of our WTAM neurons, where attributes are either part of a bundle or they are not.
	The sigmoidal activation function we use is defined as:

	\[g(x) = \frac{1}{1 + e^{-k(x - \theta)}}\]

	Where $k$ is the slope of the sigmoid, and $\theta$ is the offset.

	This allows the connection strength of our WTAM neurons to only increase when both the attribute and memory neurons are strongly activated, effectively thresholding the Hebbian learning.

	Another mechanism to increase the discreteness of our memory neurons is the use of bistable neurons.
	These neurons have two stable states, namely an upper and lower attractor state, and they can switch between these states depending on the input they receive.
	This is opposed to the leaky neurons we used before, which only have one stable state: their baseline state, to which they return over time without any external input.
	This further supports the idea that attributes are either part of a bundle or they are not, and that a bundle should either be activated or not.
	We use an extended version of our decay equation to account for the bistable nature of these neurons:

	\[\Delta r_i = -c d_{r_i} (r_i - r_L) (r_i - r_T) (r_i - r_U)\]
	
	Where $r_L$ is the lower attractor state (comparable to baseline), $r_T$ is the transition point, and $r_U$ is the upper attractor state.
	The transition point is defined as the average of the attractor states, meaning that the neuron will switch between these states when it crosses the midpoint between them.
	The decay constant $c$ is modelled to ensure that the meaning of the half-life is comparable between normal leaky neurons and bistable neurons, although it is necessarily an approximation because the decay rate depends on how close the state is to the transition point.

	By using bistable neurons, we ensure that the memory neurons are more stable and discrete, leading to a more pronounced winner-take-all dynamic where only one memory neuron is activated at any given time.

	There is an intrinsic trade-off between the effectiveness of memory allocation and memory retrieval in our WTAM neurons, which is determined by the strength of the lateral inhibition between them.
	High lateral inhibition makes the network more sensitive to noise, because only a small difference in one neuron can lead to the suppression of all others.
	On the other hand, low lateral inhibition makes the network slow to converge when selecting memory neurons through competition, leading to problems especially in cases where a novel memory bundle is being formed.
	To resolve this trade-off, we introduce FHP in the inhibitory connections between WTAM neurons.
	The lateral inhibition between memory neurons will start at a low level, reducing interference with memory retrieval by limiting the effect noise has in the short run.
	However, to prevent the competition process from becoming too slow and potentially not allocating new memory slots, the lateral inhibition will increase when neurons are consistently activated together, making it more likely that only one of them will be selected in the end.
	This adaptive lateral inhibition allows the trade-off between memory allocation and retrieval to be resolved without compromising on the performance of either.

	\subsection{Control System}

	While a winner-take-all mechanism is sufficient to assign novel combinations of attributes to Bundle memories, there are several other operations that are required to explain a bigger set of S3R phenomena.
	To this end, we add a control system (\autoref{fig:rcbm}, right top) to our model so that it can distinguish between novel and existing memories, dynamically allocate new memory slots while protecting existing ones, detecting when there are multiple ambiguous candidates, and suppress the whole system during other activity.
	We implement this control system using a set of specialized neurons that interact with the memory system, and that are responsible for activating the appropriate memory neurons and suppressing the memory system when necessary.
	We will discuss these control neurons in more detail in the following sections.
	
	One of the key operations that the control system needs to perform is novelty detection, which allows the network to selectively boost unassigned memory neurons when attributes are activated that are not associated with any existing bundles.
	The novelty detection neuron (\autoref{fig:rcbm}, control system, N) is a normal leaky neuron with slightly different parameters, chosen such that, once a memory neuron is assigned and turns from a \enquote{novel} Bundle into an existing one, it will no longer be activated by (or itself activate) the novelty detection neuron.

	In practice, this means that when a novel combination of attributes is activated, since no WTAM neuron is yet associated with these attributes, the novelty detection neuron will be activated, which in turn will boost the activation of the unassigned WTAM neurons, allowing one of them to be selected rather than a neuron that is already associated with another set of attributes.

	The other side of the coin is existing memory detection, which allows the network to boost assigned memory neurons to promote competition between them, until one is eventually selected.
	The existing memory detection neuron (\autoref{fig:rcbm}, control system, E) is a normal leaky neuron with the same properties as the novelty neuron.
	The memory neurons are connected to the existing memory detection neuron through sigmoidal synapses, meaning that only activation above a certain threshold is propagated.

	In practice, as a WTAM neuron is bound to a set of attributes, it will also activate and bind to the existing memory detection neuron.
	When, subsequently, another set of attributes is activated that is associated with the same WTAM neuron, the existing memory detection neuron will be activated, which in turn will boost the activation of the WTAM neurons, keeping unassigned WTAM neurons from being activated.

	Since the memory system heavily interacts with the semantic system, it may also interfere with other cognitive processes that require the activation of the semantic system without immediate binding or activation of associated bundles.
	To this end, we add a memory suppression neuron (\autoref{fig:rcbm}, control system, S) to our control system, which turns off all WTAM neurons, allowing for cortical activity without immediate binding or activation of associated bundles.
	This same mechanism can also be used to turn off currently active Bundle memory, for example when a new sentence starts or the currently active memory is no longer relevant.
	The inhibition here uses facilitatory FHP, still allowing some activation to come through when the network is already silent, while retaining the ability to silence very strong network activity.

	All control neurons have lateral inhibition with each other, meaning that only one state can be active at any given time.
	This makes the control system function as a sort of Markov chain with certain transition pathways between the different states.
	In particular, novel activation should always automatically transition to existing, as the connections activating the novelty detection neuron get slowly suppressed while the connections activating the existing memory detection neuron get facilitated.

	\subsection{Ambiguity Detection}
	
	Given these control neurons, it turns out that the model can already explain a large part of the S3R phenomena.
	However, it struggles when faced with input with overlapping features, since this leads to ambiguity in the retrieval process.
	This retrieval ambiguity is difficult because the memory neurons are part of a winner-take-all circuit with noise, making them very eager to converge on a single memory neuron even when the evidence is fully ambiguous.
	Reducing the strength of lateral inhibition improved this, but directly reduced the network's ability to assign memory neurons to novel bundles of attributes, as this relies on the same noise-driven competition mechanism.
	To explain this phenomenon, we implement a mechanism to detect ambiguity and suppress the winner-take-all mechanism selectively when the network is presented with ambiguous input, but not when presented with novel input.
	For instance, when a novel input like ``big'' is first introduced, the winner-take-all mechanism engages to store the relevant memory trace to a single bundle, but if a ``big cat'' and ``big dog'' are already in memory, the network must recognize the retrieval ambiguity caused by ``big'' and temporarily suppress the winner-take-all mechanism to allow both memory traces to be activated simultaneously.

	To detect ambiguity, we add a secondary pool of Multiple Activation Memory (MAM) neurons (\autoref{fig:rcbm}, bottom right), which bind in parallel with the actual memory neurons and have a 1-on-1 mapping between them, but do not have lateral inhibition so that multiple memory neurons can be activated at once.
	The mapping is created by sigmoidal synapses connecting each WTAM neuron with a corresponding MAM neuron in this secondary pool, while inhibiting all other neurons in this pool — in other words, when the WTAM neurons converge, this will also lead to convergence in the MAM neuron pool.

	Though they have no lateral inhibition between them, the MAM neurons are still suppressed by the memory suppression neuron, using the same parameters as those suppressing the WTAM neurons.
	By implementing this secondary memory pool, each bundle memory now consists of two neurons: one in the primary WTAM pool and one in the secondary MAM pool.
	When the WTAM neurons converge on a single memory neuron, the corresponding MAM neuron will also be activated, but when the WTAM neurons are ambiguous, multiple MAM neurons can be activated.

	To allow the MAM neurons to be activated directly and independently from the WTAM neurons, they have their own facilitating synapses coming from the attribute neurons.

	These connections have the same parameters as those from the attribute neurons to the WTAM neurons, except with a higher scaling factor.
	This means that, when a set of attributes is activated that is associated with one or more memories, the corresponding MAM neurons will directly be activated.
	Note that there are no reciprocal connections from the MAM neurons to the attribute neurons, as there are from the WTAM neurons, meaning that the MAM neurons are not able to activate the attributes they are associated with.
	This is necessary to prevent mixing up attributes between different memory neurons: since multiple MAM neurons can be activated at once, this would lead to the activation of multiple sets of attributes, which would then become indistinguishable at the moment of retrieval and, in the worst case, lead to the overwriting of both bundles with a mix of their attributes.
	By having these asymmetric facilitating synapses between the MAM neurons and the attribute neurons, we allow for the identification of multiple (ambiguous) memory bundles without automatically retrieving them.

	In order to detect ambiguity, we introduce an ambiguity detection neuron (\autoref{fig:rcbm}, control system, A) in the control system, which detects the simultaneous activation of two or more ambiguous bundles.
	Neurons in the MAM pool feed into this ambiguity detection neuron, whose threshold is set such that it requires multiple active MAM neurons to be activated.
	This neuron only remains active under the presence of enough input from the MAM pool, and otherwise transitions to the existing or novel control state when disambiguating evidence is found.

	The ambiguity detection neuron inhibits the WTAM neuron pool, thus postponing the resolution of the winner-take-all mechanism until disambiguating information is provided.

	In order for this ambiguity detection to be stable, we need competing bundles to both fully activate their corresponding MAM neurons, rather than one suppressing the other like in the WTAM pool.
	It turns out that, in itself, the level of activation of the MAM neurons varies greatly depending on the number of activated attributes, the amount of overlap, and the time between memory encoding and retrieval.
	To ensure that the MAM neurons are activated in a consistent and comparable manner, we use Hebbian learning driven by synchronization-locking, where FHP only takes place when the rates of both the pre- and post-synaptic neurons are similar.
	This mechanism is somewhat similar to auto-associative networks, e.g., Hopfield networks, although the neurons here are memory traces rather than individual attributes.
	Such rate-synchronization could be explained through spiking coincidence in lower-level neuron simulations, since synchronized spiking leads to coinciding postsynaptic potentials and thus to a higher likelihood of long-term potentiation.
	This provides a sort of oscillatory resonance that allows specifically for mutually boosting similar signals, so that ambiguous bundles fully activate their corresponding MAM neurons.

	First, we calculate the matrix of absolute differences between the activation states of all neurons in the MAM pool (excluding self-connections):

	\[\Delta R_{ij} = \left| \vec{r}_\text{i} - \vec{r}_\text{j} \right|\]

	Where $\vec{r}$ is the vector of activation states of all MAM neurons, and $i$ and $j$ are the indices of the source and target neurons, respectively.
	Next, we threshold these differences using a sigmoidal function, which when inverted gives us a matrix $S$ of synchrony values between all pairs of neurons:

	\[S_{ij} = 1 - \frac{1}{1 + e^{-k_S(\Delta R_{ij} - \theta_S)}}\]

	where $k_S$ and $\theta_S$ are the slope and offset of the sigmoid, respectively.
	These values are then used to calculate the overall FHP, which is the product of the synchrony and our previously defined FHP mechanism:

	\[\Delta W_\text{MAM} = \eta S \odot \mathcal{H}(\vec{r}_\text{source}, \vec{r}_\text{target}) \]

	Where $\eta$ is the FHP rate, as before, and the decay works the same as for other facilitating synapses.
	The $\odot$ operator denotes the element-wise product, and $\mathcal{H}$ is the Hebbian learning function defined previously.

	\subsection{Semantic Network}

	Of course, a memory system can only meaningfully store information if it is connected to a semantic network that encodes this information.
	Though the focus of this paper is on the memory system, we aimed to also provide a richer and more extensive semantic network than the one used by \citet{manohar_neural_2019}.
	An important reason for this is that we aim to not just explain a phenomenon in visual working memory, but to explain a wider range of S3R phenomena which require a more complex conceptual space.
	To this end, we implement a semantic network with both unimodal sensory areas and a central multimodal/transmodal area.
	It bears some resemblance to hub-and-spoke models \citep{patterson_hub-and-spoke_2016}, in that it consists of separate unimodal sensory areas (the spokes; see \autoref{fig:rcbm}, left) that converge on a central multimodal/transmodal area (the hub; see \autoref{fig:rcbm}, center).
	The internal structure of this latter area itself takes its inspiration from convergence zones \citep{damasio_brain_1989}, which we implement as semantic layers of increasing abstraction: the first layer connects to the sensory areas, the second to the first, and the third to the second, with each layer representing progressively more abstract concepts -- a graded organisation from multimodal to transmodal that also echoes the hierarchies of abstraction described by \citet{binder_neurobiology_2011}.
	Finally, we implement linguistic input and output layers that map to and from the semantic neurons (\autoref{fig:rcbm}, top, neurons not shown).
	All of these semantic neurons represent arbitrary concepts in an artificial conceptual space and language, though they were picked to provide a conceptual space that is rich enough to explain the S3R phenomena we defined.
	We use the terms \enquote{word} and \enquote{language} loosely here: our artificial language lacks some features that natural languages usually possess, for instance that it does not distinguish between subject and object, and that its words carry no part-of-speech information.
	Each word instead maps one-to-one onto a semantic attribute, so that what may look like nouns, adjectives, or verbs (especially in our English \enquote{translation} of our examples, using for example \enquote{cat}, \enquote{big}, or \enquote{meowing}) are in our model simply attributes at different levels of abstraction, distinguished only by their semantic content and not by any grammatical category.
	We adopt this deliberately simple artificial language because it isolates the memory and binding phenomena we are interested in, while preventing the model from having to additionally implement the complexities of natural language syntax and semantics.

	Our semantic model consists of two unimodal input layers (representing the visual and auditory modalities), each encoding two attribute classes  (e.g., color or pitch) with three mutually exclusive attributes each, very much like the semantic network used by \citet{manohar_neural_2019}.
	These are the spokes of our Hub-and-Spoke-like architecture, and they are connected to a concrete-semantic hidden layer, which forms the first layer of the semantic hub.
	This layer contains eight attribute neurons per unimodal area, with four neurons per attribute: one for the attribute itself and three for the properties.
	The second layer of the semantic hub connects to the first, and contains eight attribute neurons with combinations of concrete attributes, i.e., multimodal concepts like dog (e.g., the shape and sound of a prototypical dog).
	The third layer of the semantic hub connects to the second, and contains four attribute neurons with combinations of multimodal attributes, i.e., abstract concepts like animal (e.g., the concept encompassing dogs, spiders and birds, but not trees).
	We use linguistic input and output layers of 28 neurons each, the same as the total number of attribute neurons, mapping to and from these attribute neurons.

	This makes for a relatively rich semantic network for its size, with 2 modalities and 28 attributes of different kinds, for a total of 96 neurons in the semantic system.
	It is worth emphasising that this semantic network is separate from the memory system itself: the bundle-memory system interfaces with the semantic network, but its information remains represented in the semantic network rather than being copied into the memory system.
	This distinction matters when we later relate the memory system to the cortico-hippocampal circuit (see Discussion): such areas should be understood as interfacing with the semantic system, not as constituting it \citep{duff_semantic_2020}.

	\subsection{Model Parameters}

	An overview of all model parameters is provided in \autoref{table:connection_parameters} and \autoref{table:neuron_parameters}.
	Parameters were chosen through manual tuning, starting from the parameters provided by \citet{manohar_neural_2019} and adjusting them to achieve both network stability and good performance on a wide range of S3R phenomena.
	In certain cases, we analysed the interactions between different parameters to find a good balance between them, for example in the case of the lateral inhibition strength and the noise level (prior to the implementation of the ambiguity detection mechanism).
	Overall, we aimed to keep the number of parameters low by reusing the same parameters across different parts of the model where possible, for example by using the same FHP parameters for all connections from attribute neurons to memory neurons.
	Furthermore, we aimed to avoid overly sensitive parameters that would require fine-tuning to achieve good performance, instead replacing them with mechanisms that are more robust to parameter changes, for example by using the ambiguity detection mechanism to resolve the trade-off between memory allocation and retrieval in case of ambiguity.
	This means that the current parameter set is not the only one that can explain the S3R phenomena, and that there is likely a wide range of parameters that would lead to good performance.
	Importantly, we should be explicit about what we do and do not mean by calling our model neurobiologically plausible.
	We do not claim that the specific parameter values, the number of neurons, or the precise architecture of our model match those of any particular brain circuit; the model operates in arbitrary units and is deliberately minimal.
	Instead, our claim to biological plausibility rests on the \textit{mechanisms} from which the model is built: each component relies only on neural and synaptic operations that the brain is known to possess, such as fast Hebbian plasticity, lateral inhibition, facilitating synapses, and bistable neural dynamics.
	What we aimed for, then, is not a model whose parameters fall within measured physiological ranges, but one in which every mechanism has a plausible neural implementation.
	The same caveat applies to the most conspicuous simplification of our model, its four control neurons: we do not propose that novelty, existing memory, ambiguity, and suppression are each signalled by a single neuron in the brain, but model them as single units only to show, as transparently as possible, how such control signals can be computed bottom-up from an otherwise distributed memory system (see the Discussion section on demystifying the control homunculus).
	Finally, to reduce the risk of overfitting to specific items, we evaluated the model on a large dataset of randomly generated items, rather than manually selected ones.
	As such, we have shown that the current parameter set is able to explain a wide range of S3R phenomena in a robust manner.
	However, while there is no overfitting to the semantics in our dataset, it is possible and even likely that the parameters are overfitted to the syntax of our specific S3R phenomena, and future work should aim to evaluate the model on a wider range of syntactic structures to ensure its generality.

	\begin{table*}
	\renewcommand\cellalign{tl}
	\renewcommand{\arraystretch}{2}
	\begin{tabular}{lllllllll}
	\hline
	\textbf{}		& \textbf{Sensory}											& \textbf{Semantic} 																				& \textbf{WTAM} 																						& \textbf{MAM} 																						& \textbf{Amb.} 							& \textbf{Novelty} 																						& \textbf{Supp.} 							& \textbf{Ex.} \\
	\hline							
	\textbf{Sens.}	& \makecell[l]{$\gamma=-.003$\\$w_0=1$\\}					& \makecell[l]{$\gamma\in[.01,.03]$\\$w_0\in\{0,1\}$\\}												&  																										&  																									&  											&  																										&  											& \\
	\textbf{Sem.}	& \makecell[l]{$\gamma\in[.002,.01]$\\$w_0\in\{0,1\}$\\}	& \makecell[l]{$\gamma\in[-6e^{-4},.01]$\\$w_0\in[0,1]$\\}											& \makecell[l]{$\gamma=.01$\\$w_0=.2$\\$d_w=2e^{-5}$\\$\eta=.5$\\$k_\eta=50$\\$\theta_\eta=.7$\\} 		& \makecell[l]{$\gamma=.02$\\$w_0=.2$\\$d_w=2e^{-5}$\\$\eta=.5$\\$k_\eta=50$\\$\theta_\eta=.7$\\}	&  											&  																										&  											& \\
	\textbf{WTAM}	& 															& \makecell[l]{$\gamma=.02$\\$w_0=0$\\$d_w=2e^{-5}$\\$\eta=.5$\\$k_\eta=50$\\$\theta_\eta=.7$\\} 	& \makecell[l]{$\gamma=-.2$\\$w_0=.1$\\$d_w=.07$\\$\eta=.1$\\} 											& \makecell[l]{$\gamma=.05$\\$w_0\in\{-.75,1\}$\\$k_\eta=50$\\$\theta_\eta=.75$\\} 					&  											& \makecell[l]{$\gamma=.08$\\$w_0=1$\\$d_w=2e^{-5}$\\$\eta=-.005$\\$k_\eta=50$\\$\theta_\eta=.7$\\} 	&  											& \makecell[l]{$\gamma=.04$\\$w_0=1$\\$k_\eta=20$\\$\theta_\eta=.6$\\} \\
	\textbf{MAM}	&  															&																									&  																										& \makecell[l]{$\gamma=.03$\\$w_0=0$\\$d_w=.07$\\$\eta=.1$\\$k_\eta=50$\\$\theta_\eta=.3$\\} 		& \makecell[l]{$\gamma=.5$\\$w_0=1$\\} 		&  																										&  											& \\
	\textbf{Amb.}	&  															&																									& \makecell[l]{$\gamma=-.02$\\$w_0=0$\\$d_w=.07$\\$\eta=.1$\\} 											&  																									&  											& \makecell[l]{$\gamma=-.02$\\$w_0=1$\\} 																& \makecell[l]{$\gamma=-.02$\\$w_0=1$\\} 	& \makecell[l]{$\gamma=-.02$\\$w_0=1$\\} \\
	\textbf{Nov.}	&  															&																									& \makecell[l]{$\gamma=.025$\\$w_0=1$\\$d_w=2e^{-5}$\\$\eta=-.005$\\$k_\eta=50$\\$\theta_\eta=.7$\\}	&  																									& \makecell[l]{$\gamma=-.02$\\$w_0=1$\\} 	&  																										& \makecell[l]{$\gamma=-.02$\\$w_0=1$\\} 	& \makecell[l]{$\gamma=-.03$\\$w_0=1$\\} \\
	\textbf{Supp.}	&  															&																									& \makecell[l]{$\gamma=-.2$\\$w_0=0$\\$d_w=.07$\\$\eta=.1$\\} 											& \makecell[l]{$\gamma=-.2$\\$w_0=0$\\$d_w=.07$\\$\eta=.1$\\} 										& \makecell[l]{$\gamma=-.02$\\$w_0=1$\\} 	& \makecell[l]{$\gamma=-.03$\\$w_0=1$\\} 																&  											& \makecell[l]{$\gamma=-.03$\\$w_0=1$\\} \\
	\textbf{Ex.}	&  															&																									& \makecell[l]{$\gamma=.02$\\$w_0=0$\\$d_w=2e^{-5}$\\$\eta=.01$\\} 										&  																									& \makecell[l]{$\gamma=-.02$\\$w_0=1$\\} 	& \makecell[l]{$\gamma=-.02$\\$w_0=1$\\} 																& \makecell[l]{$\gamma=-.02$\\$w_0=1$\\} 	& \\
	\hline
	\end{tabular}
	\tablenote{\small Synaptic parameters for each neuron population pair. Each row represents a population of neurons in the model, indicated by the leftmost column, and displays the relevant synaptic parameters for connections to other populations, as indicated by the different columns. The different parameters are: $\gamma=$signal propagation scaling factor, $w_0=$ baseline connection strength, $\eta$ = FHP rate, $d_w$ = weight decay rate, $k_\eta$ = FHP sigmoid slope, $\theta_\eta$ = FHP sigmoid offset.}\label{table:connection_parameters}
	\end{table*}

	\begin{table*}
	\begin{tabular}{lllll}
	\hline
	\textbf{Sensory Neurons}	& \textbf{Semantic Neurons} & \textbf{WTAM Neurons} & \textbf{MAM Neurons} 	& \textbf{Control Neurons} \\
	\hline
	$\zeta=0$					& $\zeta=0$ 				& $\zeta=.005$ 			& $\zeta=0$ 			& $\zeta=0$ \\
	$b=.2$						& $b=.2$ 					& $b=.2$ 				& $b=0$ 				& $b=0$ \\
	$d_r=.034$					& $d_r=.034$ 				& $d_r=.034$			& $d_r=.034$ 			& $d_r=.034$ \\
	\hline
	\hline
	\end{tabular}
	\tablenote{\small Parameters per neuron population. Each column represents a population of neurons in the model and displays the relevant neural parameters for that population. The different parameters are: $\zeta$ = noise rate, $b$ = baseline activation, $d_r$ = firing rate decay rate.}\label{table:neuron_parameters}
	\end{table*}

	\subsection{Task Design}

	To test the model's ability to explain a wider range of S3R phenomena, we implement an artificial language and a set of tasks that require the model to explain (many of) these phenomena.
	In total, we define 14 tasks divided over 4 categories: controlled storage, memory retrieval, interleaved memories, and semantic association.
	We evaluate the model's performance on these tasks relative to expected outputs, with the expectation that our full model should be able to perform these tasks with high accuracy.

	Each test consists of 3 to 4 trials, representing different sentences that test the same phenomenon.
	These tests could test the same phenomenon with different semantic content, or different aspects of the same phenomenon (e.g., to test the model's ability to forget over time, we could test it on different time scales).
	Each trial consists of a sequence of inputs, with each input presented for 500 timesteps.
	Each input was either a word (one of the 28 conceptual input neurons), a sensory input (one of the 12 unimodal input neurons), a full stop (activating the suppression neuron), a novelty signal (A person vs. THE person, where the former indicates novelty; activating the novelty neuron), or no signal.
	This means that there are 42 different input possibilities, which are combined in different ways to create the tests.
	The absence of any input is used to measure responses at important moments in the sequence, as the model will maintain its current state without interference from input signals.
	Each trial is repeated 25 times, and performance is measured by checking whether these responses match the expected outputs.
	In total, we ran approximately 1200 trials for our primary model.

	Our test suite was divided into four categories: controlled storage, memory retrieval, interleaved memories, and semantic association.
	Semantic association tests evaluate the model's ability to use the semantic network in combination with the memory system to explain phenomena that require the association of different concepts.
	Controlled storage tests evaluate the model's ability to store information in a controlled manner, i.e., whether memory doesn't \textit{just} store and retrieve any information, but always the \textit{right} information.
	Memory retrieval tests evaluate the model's ability to retrieve information from memory, both now and over longer periods of time.
	Multiple memories tests evaluate the model's ability to store and retrieve multiple memories at once, and to distinguish between them.

	First of all, \textbf{semantic association} contains the following tests:
	\begin{itemize}
		\item \underline{Deduced association}: assesses the model's ability to infer that a novel attribute is associated with an existing memory neuron through an existing association with another attribute that is part of that bundle, i.e., to deduce reference to a memory through a concept that has not been mentioned before in combination with that memory.
		\item \underline{Correlation inference}: assesses the model's ability to infer the presence of an attribute based on the presence of another attribute in the same memory bundle, based on an association between these two attributes that has been learned before and is stored in the memory system.
		\item \underline{Sensory binding}: assesses the model's ability to indirectly bind sensory input to a memory representation, i.e., to use sensory input to activate a conceptual attribute that is then stored in memory, and vice versa to again activate the sensory representation from the memory representation.
	\end{itemize}

	Secondly, \textbf{controlled storage} contains the following tests:
	\begin{itemize}
		\item \underline{Correlation violation}: assesses the model's ability to store anti-correlated attributes in memory neurons, i.e., to combine attributes that do not normally occur together without interference.
		\item \underline{Inferred novelty}: evaluates the model's ability to infer the novelty of an information sequence, i.e., to allocate a new memory neuron for a sequence that describes a new concept, rather than associating it with an existing memory neuron.
		\item \underline{Determined novelty}: evaluates the model's ability to create a novel memory neuron, even in cases where the information might otherwise be inferred as part of an existing memory neuron, by explicitly marking the information as novel.
	\end{itemize}

	Third, \textbf{memory retrieval} contains the following tests:
	\begin{itemize}
		\item \underline{Gradual forgetting}: assesses the model's ability to forget attributes of a memory over sufficiently long periods of time.
		\item \underline{Long-term remembering}: assesses the model's ability to retain information for a certain amount of time.
		\item \underline{Memory maintenance}: evaluates the model's basic capability to retain a coherent memory from uncorrelated attributes, i.e., to implement binding and one-shot learning.
		\item \underline{Pattern completion}: measures the model's ability to complete a pattern based on partial input of a previously learned sequence, i.e., content-addressable memory.
	\end{itemize}

	Finally, \textbf{interleaved memories} contains the following tests:
	\begin{itemize}
		\item \underline{Problem of two}: assesses the model's ability to distinguish between two memory neurons with overlapping attributes, i.e., to handle competition and resolve to choose one memory neuron over another.
		\item \underline{Discourse incrementation}: assesses the model's capability for incremental binding of new information to existing memory neurons over multiple sentences, i.e., to build up an integrated memory representation over time.
		\item \underline{Many memories}: assesses the model's capability to create and retrieve up to four different memory bundles in various sequence patterns, without interference.
		\item \underline{Pattern completion with two referents}: measures the model's ability to complete a pattern based on partial input of a previously learned sequence when presented with another distractor sequence in between.
	\end{itemize}

	\subsection{Lesioning the Model}

	To investigate which connections and mechanisms are causally required for the model's functioning, we rerun each test for lesioned versions of the network.
	These lesions are introduced by not including certain connections in the network, while leaving the neurons themselves and other connections intact.
	This allows for a clean side-by-side comparison of how the model functions without some of its connections.
	We define lesions to different parts of the network:
	\begin{itemize}
		\item Existing: removes the connections to and from the existing memory detection control neuron.
		\item Ambiguity: removes the connections to and from the ambiguity detection control neuron.
		\item MAM:\ removes the connections between the WTAM neurons and the MAM neurons, as well as the connections from the MAM neurons to the attribute neurons.
		\item MAM synchrony: removes the lateral facilitatory connections between the MAM neurons that drive the synchrony detection mechanism.
		\item Sensory grounding: removes the connections between semantic layers, leading to loss of semantic information.
		\item WTAM completition: removes the lateral inhibitory connections between the WTAM neurons that drive the winner-take-all mechanism.
		\item Suppression: removes the connections to and from the memory system suppression control neuron.
		\item Novelty: removes the connections to and from the novelty detection control neuron.
		\item WTAM:\ removes the connections between the attribute neurons and the WTAM neurons, in both directions.
	\end{itemize}
	We rerun each test for each lesioned version of the network besides the full unlesioned model, for a total of around 12000 trials across all tests and lesions.

	\subsection{Tools}
	The RCBM model was implemented in Python 3.10.9 \citep{python}, using Aesara 2.9.3 \citep{aesara} for optimizing and running our simulations; NumPy 1.26.4 \citep{numpy} and Pandas 1.5.2 \citep{mckinney_data_2010} for data handling and aggregation; and Matplotlib 3.6.2 \citep{matplotlib} for data visualization.
	Analyses of collected simulation data were conducted in R 4.2.1 \citep{r-project}, using Tidyverse 2.0.0 \citep{tidyverse} for data processing, analysis, and visualization.

	\section{Results}\label{sec:results}

	\begin{figure*}
		\centering
		\includegraphics[width=\textwidth]{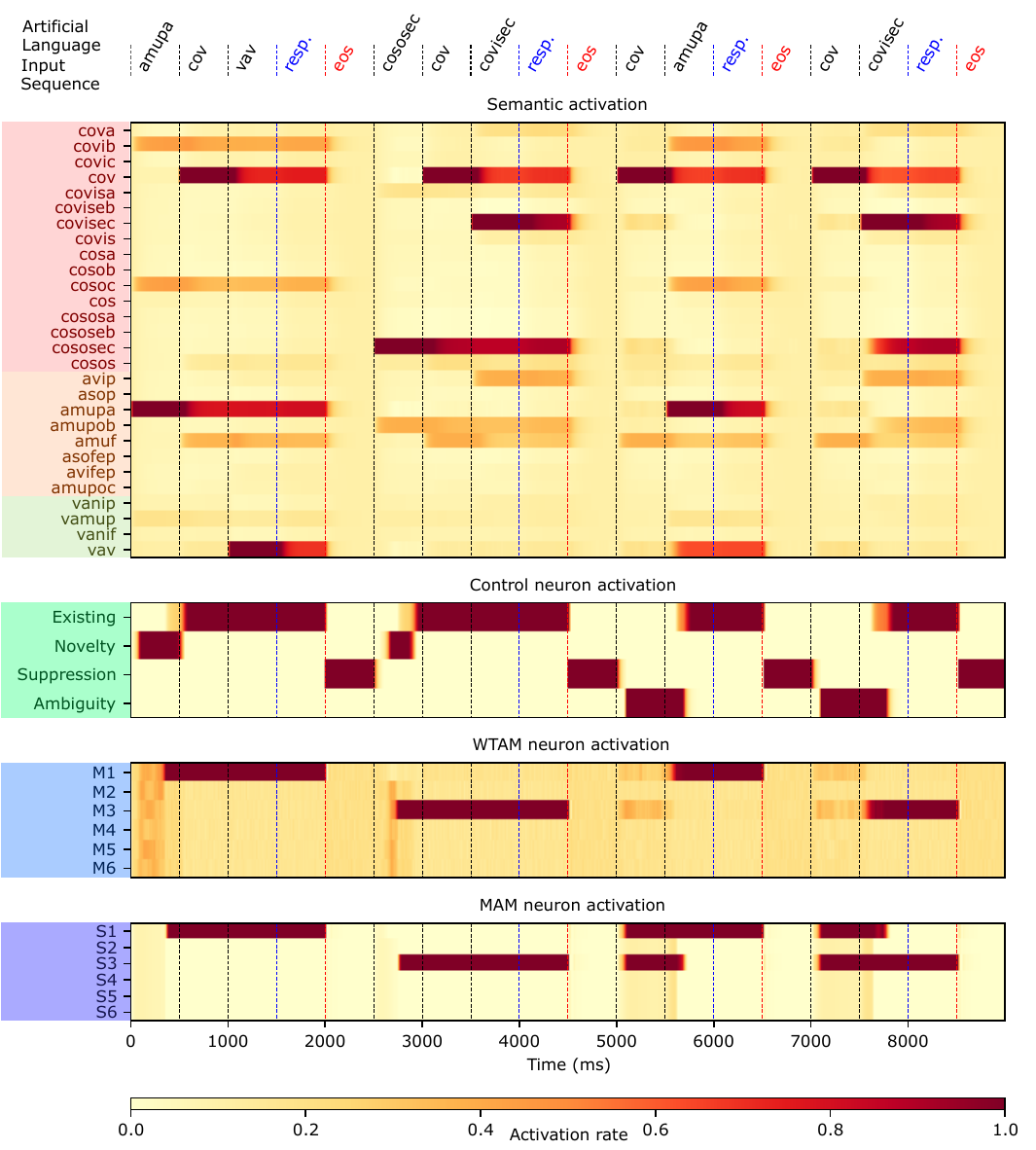}
		\caption{
			\small
			An example trial of the problem of two test with corresponding neuronal activations of the model.
			In contrast to \autoref{fig:rcbm_sequence} which shows a schematic trial of a simpler test case, here we show the full timecourse of a more complex trial.
			At the top, the trial is presented as a sequence of words in an artificial language, where each word in black corresponds to a specific semantic attribute, ``resp.'' indicates the moment of response, and ``eos'' indicates the end of a sentence.
			The panels show the activation of the semantic attribute neurons (top, red, yellow, and light green), the control system neurons (middle, green), and the Winner-Take-All Memory (WTAM, blue) and Multiple Activation Memory (MAM, purple) neurons (bottom two panels).
			These neurons and their color coding correspond to the different components of the model's architecture as seen in \autoref{fig:rcbm}.
			In the first two sentences, the model is introduced to two entities with an overlapping feature, ``cov'',
			e.g. a big blue square and a small blue triangle.
			The next two sentences each start with this overlapping feature (the blue one) and then introduce a distinguishing feature (either big or small), requiring the model to temporarily withold its response until this second, distinguishing feature is presented.
			Subsequently, the model should respond with the correct remaining attribute (square or triangle).
			The model stores the two entities in M1/S1 (sentence 1) and M3/S3 (sentence 2) in the WTAM and MAM pools, respectively (bottom);
			when presented with the overlapping feature (sentence 3/4), the ambiguity detection control neuron is activated (middle), supressing the winner-takes-all process until disambiguating information is presented.
		}\label{fig:trial_example_problem_of_two}
	\end{figure*}

	To evaluate the capabilities of the Rate-Coding Bundle Memory (RCBM) model, we designed a comprehensive set of tasks aimed at addressing key challenges in cognitive modeling, particularly those related to Symbol Recombination, Retention, and Resolution (S3R).
	These tasks were carefully constructed to test the model's ability to store, retrieve, and manipulate information in a controlled manner in a way that is relevant to how humans construct memories based on linguistic input \citep{seuren_logic_2009,baggio_balance_2011,mcelree_memory_2003}.
	Specifically, we assessed the model's performance across four categories: controlled storage, memory retrieval, tracking interleaved memories, and semantic association.
	Each task was designed to probe specific cognitive phenomena, such as one-shot learning, ambiguity resolution, and incremental memory integration, which are critical for understanding higher-order cognition. 
	Additionally, we conducted lesion studies to systematically investigate the causal necessity of individual components and mechanisms within the model, providing insights into how different subsystems contribute to its overall functionality. 
	This approach allowed us to rigorously test the model's ability to explain a wide range of cognitive phenomena, and to identify the specific mechanisms that are required for performing these tasks effectively.

\begin{figure*}
		\centering
		\includegraphics[width=\textwidth]{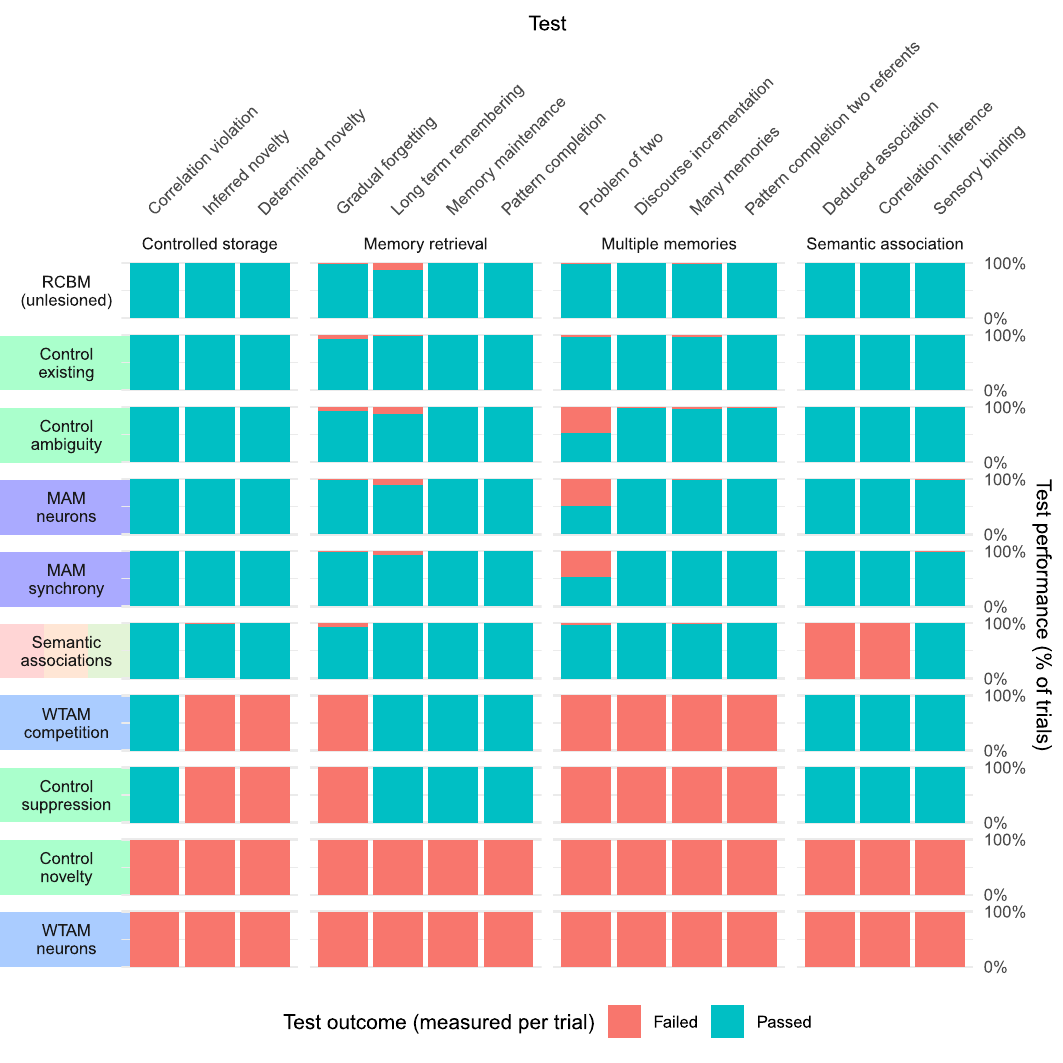}
		\caption{
			\small
			The RCBM model performs well on all tests, while lesions to the model lead to systematic reductions in performance.
			For an example of a trial of one of these tests, see \autoref{fig:trial_example_problem_of_two}.
			Each bar represents the proportion of trials (right y-axis) that the model successfully (blue) or unsuccessfully (red) completed for each test.
			The tests are displayed on the x-axis (top) and grouped into four categories.
			The different lesions are displayed on the y-axis (left), with each lesion representing a different part of the model that was turned off.
			Per test, there were up to 4 trial types that were each repeated 25 times, for a total of 12000 trials across all tests and lesions.
		}\label{fig:linguistic_lesion_results}
	\end{figure*}

	We show that the RCBM model is capable of explaining a wide range of S3R phenomena, achieving >95\% performance on all tasks (\autoref{fig:linguistic_lesion_results}, top row).
	In the category of controlled storage, the model successfully demonstrated that it could identify when a new entity was being introduced (inferred novelty), but also that it could be forced to create a new memory neuron even when the information could be inferred as part of an existing memory neuron (determined novelty).
	It also demonstrated that it could store anti-correlated attributes in memory neurons and retrieve them correctly (correlation violation).
	Under memory retrieval, the model was able to retain a coherent memory from uncorrelated attributes (memory maintenance) for prolonged periods of time (long-term remembering), and to complete a pattern based on partial input of a previously learned sequence (pattern completion).
	However, after sufficiently long periods of time, this information was forgotten (gradual forgetting), freeing up space for new information.
	All these processes worked for interleaved memories as well: the model was able to distinguish between two memory neurons with overlapping attributes (problem of two; pattern completion two referents), to incrementally bind new information to these entities over time (discourse incrementation), and to create and retrieve up to four different memory bundles in various sequence patterns (many memories).
	Finally, the model was able to use the semantic network in combination with the memory system to explain phenomena, such as inferring that a novel attribute was associated with an existing memory neuron through an existing association (deduced association), inferring the presence of an attribute based on the presence of another attribute in the same memory bundle (correlation inference), and the binding of sensory input to a memory representation (sensory binding).
	Altogether, these results demonstrate that the RCBM model is a versatile model that succeeds at its task of reliably explaining a wide range of S3R phenomena.

	In constructing the RCBM model, we demonstrate that it is possible to explain many phenomena that are symbolic in nature in a connectionist and neurobiologically plausible manner. 
	In particular, the control system of our model exemplifies this by using mechanisms based on known neural processes, such as Hebbian learning, lateral inhibition, and facilitating synapses. 
	These mechanisms allow the control system to dynamically allocate memory, detect novelty and ambiguity, suppress irrelevant information, and resolve competition between overlapping memory traces. 
	Importantly, all components of the control system are embedded within a single rate-coding framework, avoiding abstract or biologically implausible operations. 
	This implementation provides a clear example of how cognitive processes can be modeled in a way that is consistent with our current understanding of neural function.

	We redesigned the original model by \citet{manohar_neural_2019} for use in a more linguistic task by both extending the semantic network and altering the input and output layers to be sequential and symbolic, and thus more language-like, while still retaining the original more graded sensory input layers.
	By presenting pseudowords to the network one by one, the network identifies entities that are being referred to and is able to e.g., distinguish between two similar referents (\autoref{fig:trial_example_problem_of_two}).
	This requires the model to incrementally build up a coherent semantic representation of the presented information over time, which is a key aspect of language comprehension.
	In particular, here, tasks like novelty detection, the problem of two, and discourse incrementation show that the model is able to identify whether a new entity is being introduced or an existing one is being referred to and, if so, to which one.
	In addition, it shows it can incrementally bind new information to these entities over time.

	We lesioned different parts of the model to investigate which connections and mechanisms are causally required for the model's functioning.
	In doing so, we found that the model's performance was measurably reduced on particular tasks when lesions were introduced to the model (\autoref{fig:linguistic_lesion_results}), indicating that the control and memory systems are crucial for the model's ability to perform these tasks.
	We identify 4 types of impairment that we observed across the different lesions: impaired disambiguation, impaired semantic association, impaired memory capacity, and complete impairment.
	In impaired disambiguation, the model was unable to perform the problem of two task (\autoref{fig:linguistic_lesion_results}), meaning it was unable to distinguish between two memory neurons with overlapping attributes.
	This impairment occurred when the model was lesioned in the ambiguity detection control neuron, the MAM neurons, or the MAM synchrony connections (\autoref{fig:linguistic_lesion_results}); all of which are involved in the detection and resolution of ambiguity in the input.
	In impaired semantic association, the model was unable to perform the deduced association and correlation inference tasks, both of which rely on the retrieval of semantic information.
	Logically, this impairment occurred when the model was lesioned in the sensory grounding connections (\autoref{fig:linguistic_lesion_results}), which encode the semantic information in the model.
	In impaired memory capacity, the model was unable to remember more than a single memory at a time, conflating multiple memories into a single bundle and failing at a wide range of tasks.
	It occurred in two different ways: one when the model was lesioned in the WTAM competition connections (\autoref{fig:linguistic_lesion_results}), which made the model unable to select a neuron in the memory system (thus selecting them all); and one when the model was lesioned in the suppression control neuron (\autoref{fig:linguistic_lesion_results}), which made the model unable to clear its state to be able to switch between memories.
	Finally, in complete impairment, the model was unable to perform any of the tasks, indicating that the model was unable to store or retrieve any information at all.
	This occurred when the model was lesioned in the novelty control neuron or the WTAM neurons (\autoref{fig:linguistic_lesion_results}), which are both crucial for the operation of the memory system.
	Interestingly, the existing control neuron was not strictly necessary for the model to perform any of the tasks, and it seems to act like a neutral state that does not affect the model's performance when it is lesioned (\autoref{fig:linguistic_lesion_results}).

	\section{Discussion}\label{sec:discussion}

		\subsection{Summary of findings}\label{sec:summary_of_findings}

		In this study, we introduced the Rate-Coding Bundle Memory (RCBM) model, a neurobiologically plausible framework designed to address key challenges in cognitive modeling, particularly those related to Symbol Recombination, Retention, and Resolution (S3R).
		Through a series of carefully constructed tasks, we demonstrated that the RCBM model is capable of explaining a wide range of cognitive phenomena, including one-shot learning, ambiguity resolution, incremental memory integration, and pattern separation.
		These tasks required the model to dynamically store, retrieve, and manipulate information in a controlled manner, mimicking the processes involved in human memory and cognition.

		A key feature of the RCBM model is its control system, which enables the detection of novelty, ambiguity, and existing memory states, as well as the suppression of memory activity when necessary.
		This control system, implemented using mechanisms such as lateral inhibition and specialized facilitating synapses, allows the model to perform complex memory management tasks.
		By embedding all components within a single rate-coding framework, the model avoids abstract or biologically implausible operations, providing a concrete example of how symbolic-like computations can emerge from connectionist principles.

		Our results show that the RCBM model achieves near-perfect performance across a diverse set of S3R tasks, demonstrating its versatility and robustness.
		Our lesion study further revealed the causal necessity of specific components and mechanisms within the model, highlighting the importance of the control system and memory subsystems for explaining S3R phenomena.
		These findings underscore the potential of the RCBM model as a proof-of-concept of the Symbolic Subsystem Hypothesis, and provides a starting point for more comprehensive computational models of higher-order cognition.

		One limitation that is visible from our lesion results, is that our RCBM model may be over-reliant on the bundle memory subsystem.
		When the novelty control neuron is lesioned, almost everything stops working (\autoref{fig:linguistic_lesion_results}) because no new information is stored in the bundle memory subsystem anymore.
		Subsequently, our model loses all information not in bundle memory through decay, leading to loss of all information.
		In contrast, the brain also has cortical short term memory in the semantic network, which could maintain information independently of the bundle memory system \citep{sreenivasan_revisiting_2014,christophel_distributed_2017}.
		Extending our model with a more robust cortical short term memory system could be one fruitful direction to make the model less dependent on the bundle memory system.

		\subsection{Neurobiological grounding and predictions}
		We have mentioned the term \enquote{neurobiologically plausible} several times in this paper, but there is of course a difference between a mechanism being plausible and it being actually observed in the brain.
		It turns out that the RCBM model includes many mechanisms for which there is substantial empirical support in the neuroscientific literature.
		Here, we will discuss several ways in which the RCBM model is consistent with existing theories and findings in the field of neuroscience, and how properties of the RCBM model may lead to new predictions about the brain. 
		In what follows we will primarily draw on the hippocampus and dorsolateral prefrontal cortex, as these are the brain areas for which the parallels with our model are clearest, given their involvement in episodic- and working memory \citep{moscovitch_episodic_2016,curtis_persistent_2003,desposito_cognitive_2015,baddeley_working_2003}.
		However, consistent with the Symbolic SubSystem Hypothesis, we expect the functions captured by our model to be realised by a distributed subsystem spanning multiple regions -- including, as we discuss below, prefrontal, cingulate, and neuromodulatory systems -- rather than to be localised to any one structure. 
		This is precisely why we speak of a symbolic \textit{subsystem} rather than a symbolic area.
		It has recently been argued that the hippocampus and prefrontal cortex employ a comparable sequence memory algorithm \citep{whittington_tale_2025}, which supports considering these two brain regions simultaneously.

		\subsubsection{Binding mechanisms}
		
		The RCBM model can store and retrieve information through fast Hebbian plasticity. 
		It is well-established that the hippocampus is capable of fast (and long-term) Hebbian plasticity, and that this can already occur after a single presentation \citep{rolls_mechanisms_2013,kesner_ca3_2008}.
		There is more debate on the mechanisms of synaptic plasticity in the prefrontal cortex, with the dominant hypothesis being that it stores information through persistent neural firing, i.e., in the activity of the neurons themselves rather than in the synapses \citep{curtis_persistent_2003,fuster_neuron_1971}.
		However, there is increasing evidence that working memory in the dorsolateral prefrontal cortex also requires fast synaptic plasticity mechanisms, not unlike those used in the hippocampus, in particular for dealing with multi-item tasks \citep{mongillo_synaptic_2008,sreenivasan_revisiting_2014,lansner_fast_2023,stokes_activity-silent_2015,lundqvist_working_2018}.
		A third possibility is that working memory need not reside in the synapses at all, but in the intrinsic excitability of single neurons \citet{fitz_neuronal_2020}.
		The RCBM model proposes circuitry that the hippocampus and dorsolateral prefrontal cortex could implement to utilize fast Hebbian plasticity to store and retrieve information in memory.

		In addition, the RCBM model predicts that the attributes are bound to the bundle memory nodes in a conjunctive manner, meaning that multiple attributes can be bound to the same object.
		Several theories and models in the literature on the hippocampus \citep{teyler_hippocampal_2007,teyler_hippocampal_2007,oreilly_complementary_2014,schapiro_complementary_2017} and in (visual) working memory/PFC literatures \citep{manohar_neural_2019,luck_capacity_1997,wheeler_binding_2002,fougnie_object_2011}, have found evidence that attributes are indeed conjunctively bound to objects for both episodic as well as working memory.
		Our model is consistent with these findings and theories, and provides a more detailed account of how this binding process could be implemented in the brain.

		To allow storage and retrieval of the same information, the RCBM model predicts that the connections between attributes and bundle neurons must be reciprocal:
		the strengthening of an incoming connection between an attribute and a bundle should lead to the strengthening of the relevant outgoing connection as well.
		Hippocampal replay and cortical consolidation provides some indirect evidence for this idea, 
		since these phenomena can only be effective if the connections to and from the entorhinal cortex and the hippocampus are similar enough that memories encoded in the hippocampus can be successfully retrieved via the cortical-hippocampal-cortical loop \citep{siapas_coordinated_1998,maingret_hippocampo-cortical_2016,rothschild_corticalhippocampalcortical_2017}.
		Additional support for this idea is that there is topographical reciprocity: regions of the entorhinal cortex that project to particular hippocampal parts receive output from those same parts \citep{tamamaki_preservation_1995,naber_reciprocal_2001,witter_entorhinal_2021}.
		The RCBM model more strongly predicts that such reciprocal connections are necessary for the storage and retrieval of information in the brain, and that this is a general mechanism that can be used in other brain areas as well.
		
		The RCBM model predicts selective strengthening of some connections (the attributes of the same memory) and not others (the attributes of an interfering memory).
		This can be done in models with compartmentalized synapses, where the strengthening of one connection does not necessarily lead to the strengthening of all other connections of the same neuron: instead synaptic plasticity can be localized to individual dendritic spines \citep{yuste_dendritic_1995,sabatini_life_2002}.
		Indeed, there is evidence that synapses can be compartmentalized in the hippocampus \citep{yuste_dendritic_1995,matsuzaki_structural_2004}, and that compartmentalized synapses are important for encoding memories there \citep{govindarajan_dendritic_2011,yang_spine_2008}.
		In fact, the simplifications that many computational neuron models make limit the range of computations that they are able to perform, not just in the hippocampus but throughout the brain \citep{poirazi_pyramidal_2003,morita_possible_2008,legenstein_branch-specific_2011,caze_passive_2013}.
		Nevertheless, more evidence is required to confirm the prediction that the memory-related compartmentalized synapses in the hippocampus (and elsewhere) do indeed bind semantic attributes only to certain memory neurons and not others.

		Finally, a major computational advantage of bundle memory is its ability to do dynamic variable binding.
		Dynamic variable binding is an important principle of symbolic computation (see Discussion Section on why Bundle Memory is symbolic).
		Like variables in a computer program, hippocampal place cells only have stable representations within particular contexts:
		When the context changes, the representations dynamically change or remap and place cells that were active in one environment, become silent in another, while different cells become active \citep{fenton_remapping_2024,jeffery_place_2011}.
		Moreover, similar to how a variable can represent a variety of different values, the hippocampus is capable of binding a variety of different (grid cell) features to individual (place) cells \citep{jeffery_place_2011,fenton_remapping_2024}.
		Place cells can even concurrently store multiple distinct locations in a context-dependent manner \citep{park_ensemble_2011}, suggesting memories can be multiplexed, something that is not implemented in the RCBM model.
		In the context of language comprehension, hippocampal patients have been shown to have difficulty with pronoun resolution, which requires the binding of a pronoun to its referent \citep{kurczek_hippocampal_2013, covington_effect_2020}.
		Moreover, \citet{dijksterhuis_pronouns_2024} recently observed that hippocampal cells that were selective to a particular noun were later reactivated by pronouns that refer to the cells preferred noun, suggesting that the hippocampus is involved in rapidly binding pronouns to their referents, a computation that is closely related to dynamic variable binding.
		Thus, the flexible binding mechanism of the hippocampus makes it ideally suited for symbolic computation in the process of integrating semantic information into structured representations over time \citep{kazanina_neural_2023,kurth-nelson_replay_2023}.
		Importantly, referential and discourse-level integration is not exclusively conducted in the hippocampus: during sentence comprehension, the right and left inferior frontal gyri and the left angular gyrus show differential sensitivity to discourse context and world knowledge \citep{menenti_when_2009}, indicating that the binding of meaning across a discourse also draws on frontal and parietal cortex.
		Note that we use \enquote{structured} here in a specific and limited sense: semantic information organised into discrete bundles of co-bound attributes (i.e.\ objects), rather than the hierarchical, relational structure exemplified by a syntactic tree.
		As we discuss below (see the section on Hierarchy), the present model captures the former kind of structure but not yet the latter.

		\subsubsection{Distinction between the WTAM and MAM pools}

		We likewise see parallels between our model and the hippocampal circuit when it comes to the distinction between Winner-Takes-All Memory (WTAM) and Multiple Activation Memory (MAM) pools.
		More specifically, the functional distinction between WTAM (separating memories through lateral inhibition) and MAM (keeping multiple memories active through auto-association) can be seen as similar to the distinction between neuronal properties of the dentate gyrus and CA3 in the hippocampal circuit.

		In order to prevent interference between memories, which may have overlapping or contrasting attributes, the WTAM pool allows activation of only a single bundle.
		Similarly, the dentate gyrus has very sparce coding, meaning that only a small number of dentate gyrus granule cells are active during encoding or retrieval of a particular memory \citep{diamantaki_sparse_2016,neunuebel_spatial_2012,liu_optogenetic_2012}.
		To achieve activation of only a single bundle, the RCBM model implements a winner-takes-all mechanism in the the WTAM pool, which uses lateral inhibition between the different bundles.
		The dentate gyrus, one of the main gateways between entorhinal cortex and hippocampus, is known to have exceptionally strong lateral inhibition \citep{cayco-gajic_re-evaluating_2019,espinoza_parvalbumin_2018}.
		This same mechanism is used when encoding new information into memory, as the lateral inhibition between unbound WTAM neurons allows the model to allocate a single random new bundle when presented with a new item \citep{manohar_neural_2019,rolls_theory_2024}.
		This process of competition between memories also provides a decision mechanism when multiple different candidates are present, as it allows the model to accentuate the differences (=pattern separation) and select the most relevant candidate for the current context.
		Evidence from multiple sources have implicated the dentate gyrus in such pattern separation (see \citet{schmidt_disambiguating_2012} for a review).
		
		However, the WTA mechanism runs into problems when competing candidates are so similar that selecting one over the other would be too hasty.
		In these cases, a WTA mechanism by itself would force the model to arbitrarily pick one candidate over the other, even when they are equally relevant.
		Instead, we would like the model to be able to keep both candidates active at once, and to suspend the WTA mechanism until the model has gained enough information to select one of them.
		In order to solve this problem, we implemented a second pool of memory neurons, the MAM pool, which is activated in parallel with the WTAM pool.
		This MAM pool is connected to the WTAM pool in a 1-on-1 mapping, meaning that each WTAM-MAM pair stores a single bundle memory.
		Instead of lateral inhibition, the MAM pool has facilitating synchrony-based synapses, which allow for the parallel activation of multiple similar memories.
		We view these synchrony-based facilitating synapses as a rate-based version of spike-coincidence boosting mechanisms that can be found in (hippocampal) neurons \citep{bi_synaptic_1998} and spiking neural networks \citep{izhikevich_spike-timing_2004}.
		With these lateral synchrony-based facilitating synapses, the MAM pool implements a form of auto-associative memory \citep{izhikevich_spike-timing_2004,hopfield_neural_1982}, but rather than storing features of the same pattern as is typically done with auto-associative Hopfield networks, the MAM pool instead stores multiple previously activated bundle memories.
		The CA3, unlike the dentate gyrus (DG), has many excitatory recurrent connections that bind to one another, and is therefore (at the network level) thought to implement a form of auto-associative memory \citep{rolls_theory_2024,mcnaughton_hippocampal_1987}.

		By implementing both the WTAM and MAM pools, the RCBM model is able to keep multiple similar memories active at once without interference of the semantic content of the respective memories.
		Based on the above parallels, we predict that the hippocampal DG and CA3 exhibit a similar division of labour as our WTAM and MAM pools.
		We frame this as a prediction about the hippocampus because it offers the clearest test case, but the same division of labour could in principle be realised by analogous circuits elsewhere, for instance in prefrontal cortex.

		\subsubsection{Controlled Sequential retrieval}

		As a result of the aforementioned winner-take-all mechanism, the RCBM model retrieves only one bundle at a time.
		Thus, the RCBM model predicts sequential retrieval of individual bundles.
		Previous theoretical and empirical work strongly supports the idea that the hippocampus encodes and replays episodic memories in a sequential manner \citep{buzsaki_space_2018,olafsdottir_role_2018}.
		For instance, when navigating a maze, rats are known to sequentially retrieve the memories of the different paths they have taken (or will take) \citep{takahashi_episodic-like_2015,foster_reverse_2006}.
		There is also evidence suggesting that working memory in the dlPFC similarly involves sequential memory operations \citep{miller_working_2018,lisman_theta-gamma_2013}.
		
		In order to control the sequential retrieval, the model requires a mechanism to switch between active bundles.
		One way the RCBM model switches between bundles is through the suppression control neuron.
		This neuron is responsible for suppressing the activity of the current WTAM and MAM neurons, allowing the model to re-initialize competition and activate a different bundle.
		Currently, the RCBM model relies on new sensory input to trigger the retrieval of a different bundle after suppression of the previous one.

		But the RCBM model requires more than just suppression to control retrieval, since merely suppressing the current bundle does not afford much control.
		Right now, the additional control signals are primarily the novelty and ambiguity signals.
		Through our lesion analysis we show that such control mechanisms are necessary for our model to perform particular symbolic computations.
		But human behaviour is complex enough that more control can be expected:
		For instance, the RCBM model does not implement a mechanism to directly activate a different bundle memory after suppression, which could be useful for actively searching through or iterating over bundles.

		Though we expect our control system to be greatly simplified compared to the actual brain, we believe that the model's control states are used in the brain, and that they are necessary for the brain to perform accurate retrieval, especially in problem of two cases, which requires the detection of ambiguity.
		In particular, we predict that the brain should be able to detect novelty and ambiguity in the input, and to use this information to control the memory system.
		Given this prediction, it should be possible to read out these control variables from the brain, using neuroimaging techniques such as fMRI or EEG\@.
	
		\subsubsection{Localization of control signals}
		There already exists a wealth of neuroimaging evidence reporting novelty and ambiguity signals in the brain.
		Event related potentials (ERPs) such as the N400 and the Late Positive Complex (LPC/P300) are sensitive to novelty \citep{kutas_n400_2009,friedman_eventrelated_2000,barry_components_2020}, and the Nref has been found to track (referential) ambiguity \citep{van_berkum_event-related_2003,van_berkum_early_1999}.
		Furthermore, a sensitivity to novelty has been observed in the medial temporal lobe, which includes the hippocampus (HPC), the dorsolateral prefrontal cortex (dlPFC), the orbitofrontal cortex (OFC), the striatum (S), and the basal forebrain (BF) \citep{kishiyama_novelty_2009,fredes_role_2021,kafkas_how_2018,zhang_surprise_2022,geiger_novelty_2018,petrides_orbitofrontal_2007}.
		Specific evidence for a sensitivity to memory retrieval ambiguity or interference is less clearcut, but brain areas like the anterior cingulate cortex (ACC), the medial prefrontal cortex (mPFC), the left inferior frontal gyrus (lIFG) and the hippocampus have all been associated to these processes \citep{badre_left_2007,jonides_mind_2008,anderson_prefrontalhippocampal_2016,van_kesteren_how_2012}.
		The above areas are logical candidates either for the control networks in our model (ACC, OFC, S, BF, mPFC, lIFG) or for the bundle memory module itself (HPC, dlPFC).

		However, detection of control signals alone is clearly not sufficient.
		The RCBM model further predicts that these ambiguity and novelty are used in the control of memory.
		The strongest evidence for a relationship between novelty signals and memory control comes from the VTA-hippocampus loop:
		The hippocampus detects newly arrived information not yet stored in memory, sends this novelty signal (indirectly) to the ventral tegmental area (VTA), which in turn sends dopamine back to the hippocampus to boost memory encoding through the effect dopamine has on long-term potentation there \citep{lisman_hippocampal-vta_2005,duszkiewicz_novelty_2019}.
		Similar mechanisms might be at play in the prefrontal cortex as well:
		For instance, the dorsolateral prefrontal cortex has been implicated in novelty-based benefits on memory \citep{kishiyama_novelty_2009,geiger_novelty_2018};
		Similarly, the orbitofrontal cortex is associated with the detection of novelty, and becomes more active during memory encoding specifically \citep{petrides_orbitofrontal_2007,rolls_theory_2024}.
		As for the relationship between ambiguity detection and memory control, the negative effects of similarity-based interference on memory encoding and retrieval are well-established \citep{anderson_interference_1996}.
		For instance, \citet{mcelree_memory_2003} found that memory for a target word was worse when it was preceded by a similar word, compared to when it was preceded by a dissimilar word.
		However, the brain can overcome similarity-based interference by applying cognitive control \citep{amer_extra-hippocampal_2023,badre_left_2007,jonides_mind_2008,anderson_prefrontalhippocampal_2016,van_kesteren_how_2012}, as in our model.
		This supports the idea that the brain is capable of actively mitigating negative effects of ambiguity to resolve interference between similar memories.
		Together, these findings suggest that the brain is capable of detecting novelty and ambiguity, and moreover that these signals are used to affect memory encoding.

		\subsubsection{Transmission of control signals}

		The release of neuromodulators such as dopamine and norepinephrine for novelty and acetylcholine for both novelty and ambiguity are good candidates for ways in which these control signals could cascade and causally influence the neurons in the memory system \citep{kafkas_how_2018,bazzari_neuromodulators_2019,palacios-filardo_neuromodulation_2019}.
		Nuclei in the midbrain are able to release these neuromodulators to many (frontal) areas at the same time, including to the hippocampus and the prefrontal cortex and they are implicated in cognitive control of behaviour \citep{aston-jones_integrative_2005,beeler_synchronicity_2019,cools_inverted-ushaped_2011,hasselmo_cholinergic_2006,zaborszky_specific_2018}.
		Release of neuromodulators can be triggered by prefrontal areas, such as the anterior cingulate cortex (ACC) and the orbitofrontal cortex (OPFC), two areas that are linked to cognitive control \citep{avery_neuromodulatory_2017}.
		As mentioned in the previous section, dopamine — released in response to novelty — is known to have an effect on memory encoding, in both the hippocampus and the dorsolateral prefrontal cortex, through the effect dopamine has on long-term potentation \citep{lisman_hippocampal-vta_2005,duszkiewicz_novelty_2019}.
		In addition, there is direct causal evidence that dopamine (and other catecholaminergic neuromodulators) affect semantic processing: pharmacologically raising catecholamine levels alters the N400 response to semantic (in)congruence during sentence comprehension \citep{tan_catecholaminergic_2020}.
		This supports the idea that neuromodulatory control signals could potentially act on \emph{both} memory and semantic processes.
		In the hippocampus, acetylcholine is able to push certain neurons into high and sustained firing rates, which is thought to be a mechanism for memory encoding \citep{hasselmo_proposed_2002,douchamps_evidence_2013}.
		Moreover, when the cholinergic system is blocked, memory encoding is impaired \citep{maurer_cholinergic_2017}.
		In addition, the cholinergic system — known for its role in attention and cognitive control — is also important for pattern separation in the hippocampus \citep{raza_hipp_2017,myers_role_2009}.
		Acetylcholine from the medial septum to the dentate gyrus (DG) in the hippocampus can increase lateral inhibition in the DG, which is thought to be an important mechanism by which pattern separation is achieved \citep{raza_hipp_2017,myers_role_2009}.
		Acetylcholine could therefore provide a mechanism to exert novelty and/or ambiguity control in the brain.

		Neuromodulators like dopamine and acetylcholine act diffusely: unspecific to particular neurons \citep{liu_spatial_2021,ozcete_mechanisms_2024}. 
		This is similar to how in our model the novelty node is connected to all memory nodes. 
		One difference is that in our model the novelty signal only has an effect on memory nodes that are unbound, as determined by the connections from the novelty node to the memory nodes. 
		If neuromodulators like dopamine and acetylcholine are indeed novelty signals and unspecific, then we would predict that bound memory nodes have an internal state that is different from unbound memory nodes, which determines whether they engage in new long-term potentiation and how they respond to the dopamine and acetylcholine signals. 
		There is some empirical evidence for such internal states: for instance, neurons have homeostatic or metaplastic mechanisms in which recent long-term potentiation can lead to inhibition of further long-term potentiation \citep{abraham_metaplasticity_2008,huang_influence_1992}.
		This could provide a concrete neurobiological mechanism for the assignment of novel memories to unbound memory nodes used in our model.
		
		Acetylcholine is also linked to sequential memory retrieval through its role in theta oscillations:
		Cholinergic (as well as GABAergic and glutamatergic) projections from the medial septum regulate the theta rhythm in the hippocampus \citep{nunez_theta_2021}. 
		One prominent hypothesis is that memory encoding takes place on the peaks of this theta rhythm, while retrieval takes place on the troughs, and that this is mediated by the cholinergic signal \citep{hasselmo_proposed_2002,douchamps_evidence_2013}. 
		On this view, acetylcholine also schedules serial memory encoding and retrieval operations, similar to how our model's novelty and existing control signals decide between designating either a new unbound memory node ($\approx$encoding) or one of the bound memory nodes ($\approx$retrieval).
		Thus, the separation of encoding and retrieval by the theta cycle could be the brain's way of implementing a form of serial scheduling of memory operations, similar to RCBM\@.
	
		To summarize, the findings above corroborate that some of the control states in the RCBM model are measurable in the brain and that they are connected to memory processes.
		The brain may propagate these control states by controlled release of certain neuromodulators known to influence memory encoding and retrieval.
		This suggests that the brain has a memory control system that uses similar control signals as those in our model to regulate memory encoding and retrieval.
		However, it should also be clear from this discussion that the control system in the RCBM model is a great simplification of the complexity of the brain's memory control system and should thus not be taken literally.
		In that same vein, our RCBM model is not a neurobiologically \emph{realistic} implementation of a symbolic subsystem, but it does provide plausible neurobiological mechanisms through which a symbolic subsystem could be implemented in the brain -- illustrated in particular by the parallels between the RCBM model and the cortico-hippocampal circuit.
		Thus, while there are many interesting parallels, there are also clear differences between the RCBM model and the brain.
		The RCBM model merely serves as a proof of concept of how a memory control system could be implemented through neurobiologically plausible mechanisms and clarifies how such mechanisms can be exploited in the service of symbolic computations.

		\subsection{What makes Bundle Memory symbolic?}\label{subsec:symbolic}

		We believe that a digital storage medium is a prerequisite for symbolic computation.
		It enables the capacity of storing discrete representations of information — variables or symbols.
		Since these representations are stored in a discrete way, they can be manipulated and later retrieved, enabling a form of read-write addressable memory \citep{atkinson_human_1968,gallistel_memory_2009,kumaran_what_2016}.
		To create a digital storage medium, we implemented mechanisms that discretize the continuous state space of neural networks, e.g., bistable neurons, winner-takes-all lateral inhibition, and sigmoidal activation functions. 
		In our model, symbols are represented by WTAM-MAM neuron pairs, which form the discrete units that we refer to as bundles.
		This bundle can then be manipulated as a single unit, i.e., as an individual object (or token) distinct from its bound attributes (or types), yet simultaneously instantiating them \citep{fodor_connectionism_1988}.
		The distinguishing of types from tokens, in turn, enables a form of indirection or abstraction where syntactic operations can be performed on the created symbols — instead of operating on the semantic content directly \citep{fodor_connectionism_1988}.
		Since the model can perform syntactic operations on these symbols without needing to know the semantic content of the attributes, these operations are independent and generalizable to newly learned attributes, which is what affords systematicity \citep{fodor_connectionism_1988}.
		Moreover, this means that the contents of the bundles can be completely arbitrary, much like the contents of a variable in a computer program.
		In essence, this arbitrary binding to discrete symbols, also known as dynamic variable binding, is what provides flexibility.
		It enables symbolic operations like fully compositional processing of information \citep{marcus_algebraic_2001,hummel_getting_2011,hummel_symbolic-connectionist_2003}, since it makes possible the binding of and control over never before seen combinations of attributes.
		The properties of discretization, read-write addressable memory, type-token distinction, syntactic operations, systematicity, dynamic variable binding and compositionality are all hallmarks of symbolic computation.

		The symbolic operations implemented in RCBM allow it to regulate its own read-write operations to perform proper memory management, which is necessary for one-shot representation of more complex sequences of information, e.g., by preventing interference coming from multiple similar items.
		While the RCBM model right now only has a limited set of symbolic operations it can perform, it is easy to imagine new types of symbolic operations that can be implemented at a later time. 
		For instance, it should be relatively straightforward to implement comparisons or logical conjunctions or disjunctions between different bundles. 
		Likewise, it is easy to imagine some kind of serial search or other forms of iteration over multiple active bundles.
		Therefore, by taking this first step of implementing discrete symbols in a neural network alongside some symbolic operations, we have opened the way to extend the RCBM model with other symbolic operations, including ones that are naturally performed by humans.

		\subsection{Demystifying the control homunculus}

		Executive control has long been a central concept in explaining a wide range of cognitive phenomena. 
		However, the concept of executive control has been criticized for delegating the solving of all hard cognitive problems to an underspecified, centralized, top-down control system, which is sometimes derogatorily labeled a ``homunculus'' \citep{monsell_control_2000}.
		We will briefly address the two distinct concerns, proposed by \citet{monsell_control_2000}, raised in the debate about the control homunculus.

		The first concern put forward by the control homunculus critique is that it often remains unclear how exactly this executive control is implemented in neural tissue.
		The control homunculus argument was levied against theories that posited the existence of a control system that could solve the hard problems for which cognitive scientists had no mechanism or concrete idea for how to implement them.
		In contrast, our model provides clear neurobiological mechanisms and circuitry for some control operations that the brain might be carrying out.
		Though we also propose potential future control mechanisms, all of our primary claims are supported by specific mechanisms that are currently implemented in the model, and it should thus be entirely clear how our current version of executive control is implemented.

		The second concern is that it is often questioned whether the idea of a centralized, top-down control system can be reconciled with the evidence that the brain is a highly distributed system.
		On the one hand it seems the controlled decisions we make are discrete and unified, e.g., about what motor action to perform or what to pay attention to, and thus seemingly require a single central arbiter.
		On the other hand, the brain is a distributed system where there does not seem to be a unitary central command:
		multiple brain areas have been implicated in control processes, such as the prefrontal cortex, the anterior cingulate cortex, the basal ganglia, the thalamus, and the posterior parietal cortex \citep{menon_role_2022,uddin_towards_2019}.
		These areas do not act in a vacuum and instead interact with and are co-dependent on yet more brain areas, such as the sensory and motor cortices, the hippocampus, the midbrain, and the amygdala \citep{pessoa_emotion_2018,yavas_interactions_2019}.
		With our model we show that there is a way out of this paradox:	control in our model has a dedicated top-down control structure that is distinct from the semantic network, abstracting away the bottom-up semantic details and enabling efficient repurposing of ``syntactic'' control operations.
		However, the control system is clearly not solving the problems by itself, since it is recurrently interacting with the semantic network.
		This recurrent interaction entails that while there is a division of labour between representing semantic information (bottom-up) and making more abstract decisions about the control of (cognitive) behaviour (top-down), this process itself is tightly coupled.
		In the bottom-up direction, the type of control exerted is strongly dependent on bottom-up input coming from the semantic network.
		For instance, if the bottom-up input is novel or ambiguous, the control system will decide to allocate a new unbound memory neuron in the bundle memory system.
		In the top-down direction, the control system can decide to inhibit the bundle retrieval process, thereby guiding if and when bottom-up information is processed, stored and retrieved.
		Thus, the control system does to some extent have centralized top-down control over the semantic network, yet their recurrent interactions imply that the two subsystems ultimately form a single cohesive system.
	
		Moreover, the control system itself should not be seen as purely centralized either.
		It may appear from our depictions of the RCBM model that the control system is implemented in only four single neurons.
		To be clear, we do not believe these control states to literally be represented by single neurons in the brain, but rather that they are states represented by a distributed network of neurons.
		But more importantly, the RCBM control system is less centralized than it appears because the four neurons cannot exert control by themselves: the detection and resolution of control states is instead largely implemented in (synaptic connections to/from) the WTAM and MAM pools instead.
		The WTAM pool automatically assigns novel bundles to unused neurons through a combination of background noise and lateral inhibition, in tandem with the novelty and existing control neurons.
		The MAM pool detects ambiguity by boosting the activity of similar memories through synchrony-based facilitating synapses, which together with the ambiguity control neuron allows the model to suppress memory retrieval until disambiguating information is provided.
		In conclusion, the control system in our model is not a homunculus that solves all hard cognitive problems by itself.
		Instead, to be able to exert control, it depends on and interacts extensively with both the semantic network as well as the WTAM and MAM pools.

		\subsection{RCBM exemplifies how dynamical systems can be symbolic}

		In part due to the critique of the control homunculus, there has been a shift in cognitive neuroscience towards dynamical systems models of the brain.
		These models assume that the brain can be understood as a distributed dynamical system that operates over state spaces, and that the computations the brain performs can be seen as trajectories over these state spaces—rather than as a symbolic system with a centralized top-down control system \citep{van_gelder_dynamical_1998,van_gelder_what_1995,clark_being_1996,buonomano_state-dependent_2009}.
		On this view, attractors represent stable states in the state space that the brain can settle into, and the transitions between these states are governed by the dynamics of the system.
		Bifurcation theory can then be used to understand how the system can transition between different attractors, thereby controlling the behaviour of the system. 
		These dynamical system models have been used to explain several cognitive phenomena, such as decision-making, semantics, and (verbal) working memory \citep{fitz_neuronal_2020,spivey_continuous_2006,wang_probabilistic_2002}.
		Hopfield networks are perhaps the most famous example of a dynamical system model that can store memories as distinct attractors and retrieve them by moving back into the attractor state of the to-be-retrieved memory \citep{hopfield_neural_1982}.
		Yet, so far it has remained difficult to reconcile dynamical system models with the symbolic nature of cognition.

		Still, in contrast to what may be believed \citep{van_gelder_what_1995}, a more abstract, symbolic interpretation of the brain is not incompatible with a dynamical systems view.
		Our model—though we can abstractly describe it as a discrete, symbolic computer that performs controlled read and write operations over states in (bundle) memory—at the same time just \textit{is} a continuous dynamical system, as we have implemented it as a system of coupled differential equations.
		If we view our model as a single dynamical system, invariant to different input sequences (which we would not advise), then we may calculate a lower bound of the number of attractor states as follows:
		first of all, each possible combination of attributes ($2^N$) can be bound to each of the available bundles ($M$), for a total of $2^{(N \times M)}$ possible memory states.
		Additionally, for each of these memory states, one of the $M$ memories can be active, or the entire system can be inactive (i.e., all neurons are silent).
		Under these assumptions, the total number of possible attractor states in our model is given by \((M + 1) \times 2^{M \times N}\).
		For \(M = 6\) memory neurons and \(N = 28\) semantic attributes, this yields approximately \(10^{51}\) possible attractor states.
\footnote{
			For more realistic values of \(M\) and \(N\), the number of possible states quickly becomes astronomical.
			For instance, if we assume \(M = 100\) and \(N = 10000\), the number of possible states is on the order of \(10^{300000}\).
		}
		In addition, one may expect many more attractor states still if we also consider the control system and, in particular, the MAM pool.
		This exemplifies the compositional properties of the symbolic subsystem of bundle memory and shows its effectiveness as a memory system \citep{sommers_bundle_nodate}.
		However, the more interesting point is not that there \textit{are} many attractor states, but rather that transitions between these states can systematically and usefully be controlled.
		For instance, when the system is presented with an attribute of an existing bundle, the activation of the correct bundle memory is part of the attractor state: settling back into that attractor state then causes the retrieval of other bound features.
		When instead the system is presented with a novel set of features, the system detects an absence of memory attractor states (=novelty detection), which in turn causes the system to form a new attractor state that stores the new features.
		While viewing the RCBM model as a dynamical system can be enlightening, we believe that an abstract description helps to provide a more understandable explanation of the transitions between all of these different states, as it allows us to describe the model in terms of cognitive operations such as controlled memory storage and retrieval.

	\subsection{Hierarchy} 

		While RCBM is capable of flat compositionality, it still lacks the capacity for hierarchical compositionality.
		This means that stereotypical symbolic operations like relational binding are still difficult for the conjunctive binding mechanism of the current model.
		Relational binding is the process of combining two or more elements into a single structured representation \citep{fodor_connectionism_1988,marcus_algebraic_2001}.
		This is especially important for sentences like ``A man bit his dog'', where the relation is not commutative, meaning that the order of words matters: in this case it signals who does what to whom.
		Previous models like LISA/DORA deal with relational binding by implementing two-place predicates as combinations of one-place predicates or roles, e.g., \texttt{Bites()} and \texttt{IsBitten()}.
		They then dynamically bind the fillers to their roles, e.g., \texttt{Bites(a man)} and \texttt{IsBitten(his dog)} (Hummel \& Holyoak, 2003; Hummel \& Holyoak, 1997; Seuren, 2009; Doumas et al., 2022)
		We can borrow this idea of using one-place predicates to \enquote{flatten} the relational binding problem into multiple bundles to allow storing structured representations.
		However, this would require binding between the bundles, which is not possible in the current model.
		We could potentially allow facilitating binding between WTAM neurons and MAM neurons, which would allow us to bind one bundle as a attribute to another bundle.
		This hierarchy of bundles would then be able to resemble the hierarchical structure found in linguistic tree structures or the graphs used to represent visual scenes. 
		Without this or a similar capacity for hierarchical compositionality, our RCBM model will struggle to understand the many sentences and visual scenes containing hierarchical, non-commutative relations. 
		The main future challenge will be to extend the control network so that it can orchestrate these hierarchical bindings. 

		\subsection{Previous work}

		Existing theories and models already employ aspects of Bundle Memory \citep{lades_distortion_1993,green_what_2021,sanger_expansion_2020,kriete_indirection_2013,meeter_tracelink_2005}.
		However, bundle memory-like models are spread across the vast literature and often do not explain the same behavioural phenomena.
		For instance, we build on work by \citet{manohar_neural_2019} who has similarly used fast hebbian learning between attributes and conjunctive (=bundle) neurons to model working memory. 
		The Dynamic Link Architecture is another example of a Bundle Memory-like model that implements dynamic on-the-fly representations based on short term synaptic plasticity \citep{lades_distortion_1993,domany_correlation_1994}.
		We build upon these existing bundle memory-like models by adding a memory control system, which enables more complex cognitive operations on these symbols than previously possible.
		
		In addition to bundle memory-like theories and models, there are other connectionist/symbolic-hybrid models that attempt to explain the same phenomena through different mechanisms than Bundle Memory.
		Vector addition can potentially accomplish dynamic binding \citep{hummel_getting_2011}, even if it requires a very specific and computationally inefficient representational format that inherently struggles with hierarchical compositionality \citep{fodor_connectionism_1988,marcus_algebraic_2001}.
		Vector symbolic architectures are another way to bind the vector activations of different objects together in various ways \citep{schlegel_comparison_2022,padilla_neurobiologically_2014,gayler_vector_2004,smolensky_tensor_1990,plate_holographic_1995}.
		These architectures are currently the state of the art in relational binding.
		One such model, Spaun, has used the idea of semantic pointers to solve the binding problem \citep{stewart_spaun_2012,eliasmith_how_2013}.
		Semantic pointers are reduced representations with little to no semantic content themselves that point to the detailed semantic content in the semantic network \citep{hinton_mapping_1990}.
		Another type of model, DORA/LISA, uses temporal synchrony, rather than synaptic binding, to dynamically bind roles to fillers \citep{doumas_theory_2008,doumas_relation_2022}.
		At a coarse-grain level, these models are similar to Bundle Memory, though differ in the details of how they implement the binding operation — sometimes yielding different behavioural predictions.
		Not every such hybrid model has a neurobiologically plausible implementation, however: for instance, the binding operations of vector-symbolic architectures are typically not considered biologically plausible.
		In some sense, one contribution of bundle memory is that it provides a neurobiologically plausible implementation of these binding mechanisms.

		In addition to these computational models, there are also several high-level theories that have proposed similar ideas.
		Object files are an existing bundle memory-like idea from the visual search literature \citep{green_what_2021,kahneman_reviewing_1992,pylyshyn_role_1989}.
		Similarly, in the hippocampus literature, indexing theory speaks of binding between contentless indexes that point to information in semantic memory as a popular theory for episodic memory in the hippocampus \citep{teyler_hippocampal_1986,teyler_hippocampal_2007}. 
		Indexing theory has recently also been extended to model working memory in PFC \citep{fiebig_indexing_2020}.
		Although the ideas of object files, indirection, pointers or indexes have a long tradition in cognitive science \citep{pylyshyn_role_1989,kahneman_reviewing_1992}, for a long time it has been largely forgotten or ignored in cognitive (computational) neuroscience.
		More recent literature has re-emphasised the need for these ideas in the domains of vision, language and working memory \citep{awh_working_2025,quilty-dunn_best_2022,green_what_2021}.
		We view the fact that this same conclusion is being drawn in a broad range of different functional behaviours as evidence that bundle-like memory models are a convergence point for many different cognitive functions, and thus that the brain is likely to implement a hybrid symbolic/connectionist system of this kind.

		\subsection{Emerging symbolic computation and LLMs}

		Connectionists have long argued that symbolic behaviour can emerge without being put in place by hand \citep{mcclelland_parallel_1986}.
		Transformer-based LLMs have finally demonstrated this: research on LLMs has shown that they are able to perform reasonably well on a wide range of cognitive tasks, including reasoning, problem solving, and even some forms of creativity \citep{wei_chain--thought_2022,brown_language_2020}.
		Moreover, they exhibit some aspects of symbolic computation like variable binding by using the residual stream as addressable memory \citep{wu_how_2025,feng_how_2024}.
		Emergence, here, may refer to the fact that a model is able to perform these tasks without being explicitly programmed to do so, but rather by learning, e.g., from large amounts of data.
		Alternatively, ``emergence'' may mean that a higher-level property of a model arises from the interactions between its lower-level components, which may not exhibit these properties on their own.
		LLMs seem to exhibit both of these types of emergence.
		Our model only has the latter: the symbolic operations emerge entirely from the connectionist wiring of the neural network, but are not the result of a learning process.
		However, our model is not intended to be capable of the former type of emergence: it is a model of how these operations can be performed in a fully developed neural network, rather than a model of how these operations can be learned.
		As such, we do not make any (strong) claims about the learning process of our model, and leave the question of whether symbolic computation emerges through evolution or through developmental processes open for future work.
		
		It should also be noted that the symbolic operations in connectionist models like LLMs do not emerge out of nowhere, but are at least in part a result of the input data used to train the model.
		Since language consists of discrete tokens/symbols, and has syntactic, compositional structure, these models must learn symbolic operations in order to be able to correctly process language.
		In that sense, all models that correctly process language must, by definition, be able to perform symbolic operations of some kind.
		This implies that at least some symbolic operations are multiple realizable.
		However, architecture does matter, and there is a reason why transformers are currently better at language processing than previous language model architectures like LSTMs or RNNs \citep{sorodoc_probing_2020,vaswani_attention_2017}.
		Unlike these other architectures, LLMs have discrete attention heads that can each attend to different parts of the input, which may allow them to function like bundles.
		For example, words that refer to some earlier entity in the text often end up in the same attention head as the entity they refer to \citep{pandit_probing_2021,clark_what_2019}.
		It could be that the ability to bundle these words together through attention is what allows LLMs to perform (certain) symbolic operations.
		
		Another potential advantage of transformer-based LLMs over previous language models is that they can process the entire input at once, rather than sequentially.
		Consequently, they require no short-term memory, as all words in the context are simultaneously available to them as input.
		This computational advantage, however, translates into a disadvantage when viewed as a model of the brain, as we know that the brain lacks this advantage: we cannot read entire book chapters at once, but rather have to read them (more or less) word for word.
		This means that we require a way of incrementally parsing and building up representations of the input to then store these representations in memory and later retrieve them when needed \cite{gernsbacher_process_1990,mcelree_memory_2003,johnson-laird_mental_1983}, which is exactly the type of phenomenon that Bundle Memory is designed to explain.
		Aside from the input, the learning trajectory of LLMs is also unlike that of humans: we do not require repeated exposure to all the information on the internet in order to learn language.
		As such, though they may provide an alternative way of performing symbolic operations, LLMs are unlikely to be a good model of how the brain performs these operations.

		\section{Conclusion}\label{sec:conclusion}

		In this work, we have presented the Rate-Coding Bundle Memory (RCBM) model as a neurobiologically plausible instantiation of the Symbolic Subsystem Hypothesis. 
		Our results demonstrate that a hybrid connectionist-symbolic architecture, grounded in neural mechanisms, can explain a wide range of cognitive phenomena that have traditionally been challenging for purely connectionist or symbolic models alone. 
		By embedding a symbolic subsystem within a fundamentally connectionist framework, RCBM bridges the gap between the brain's ability to represent continuous, graded information and its capacity for discrete, symbolic operations.

		Specifically, we showed that the RCBM model can perform one-shot learning, resolve ambiguity, incrementally integrate information, and manage multiple memory traces—key aspects of the S3R phenomena. 
		Lesion studies further revealed that these capabilities depend critically on the interplay between memory and control subsystems, highlighting the necessity of specialized mechanisms for novelty, ambiguity, and memory management. 
		Importantly, all these operations emerge from neurobiologically plausible processes such as Hebbian learning, lateral inhibition, and facilitating synapses, without recourse to abstract or biologically implausible computations.

		These findings support the central claim of the Symbolic Subsystem Hypothesis: that the brain implements a symbolic system within its connectionist substrate, enabling the emergence of discrete, manipulable symbols from continuous neural dynamics. 
		This symbolic subsystem allows the brain to flexibly recombine, retain, and resolve information, supporting the structured, compositional reasoning that characterizes human cognition. 
		At the same time, the underlying connectionist architecture ensures that meaning remains grounded, and robust to noise—reflecting the graded, embodied nature of our mental representations. 
		Thus, the scientific debate between proponents of discrete symbols and proponents of continuous, graded representational spaces need not be decided in favour of a single camp: the brain most likely uses both.

	\printbibliography[]

	\clearpage
	\section{Supplementary Materials}
	
	\begin{table*}
		\caption{Example sentences of different tests}
		\begingroup
		\def\arraystretch{1.1}
		\begin{tabularx}{\textwidth}{lX}
			\hline
			\textbf{Test} &
			\textbf{Example Sentences} \\
			\hline
			\textbf{Memory retrieval} & \\
			Memory Maintenance &
			There is a small, black, hungry cat. \textit{(Is the cat small, black and hungry?)} \\
			Pattern Completion &
			There is a black, meowing cat. ... The cat is black. \textit{(Is the black cat meowing?)}\\
			Long Term Remembering &
			There is a meowing cat. ...... \textit{(Is the cat meowing?)} \\
			Gradual Forgetting &
			There is a meowing cat. ............ \textit{(Is the cat not meowing?)} \\
			\\
			\textbf{Semantic association} & \\
			Sensory Binding &
			There is a black cat. <meowing sound> \textit{(Is the black cat meowing?)} \\
			Correlation Inference &
			The animal is meowing. \textit{(Is the animal a cat?)} \\
			Deduced Association &
			The animal is meowing. The cat is hungry. \textit{(Is the meowing animal hungry?)} \\
			\\
			\textbf{Controlled storage} & \\
			Correlation Violation &
			The dog is meowing. \textit{(Is the dog meowing?)} \\
			Inferred Novelty &
			There is a meowing cat. The plane is large. \textit{(Is the large plane neither meowing nor a cat?)} \\
			Determined Novelty &
			There is a meowing cat. There is a large plane. There is a black cat. \textit{(Is the black cat not meowing?)} There is a red plane. \textit{(Is the red plane not large?)} \\
			\\
			\textbf{Interleaved memories} & \\
			Pattern Completion Two Referents &
			There is a black meowing cat. There is a large, wooden plane. \textit{(Is the black cat meowing but not large nor wooden?)} \\
			Discourse Incrementation &
			There is a meowing cat. There is a barking dog. The cat is hungry. \textit{(Is the hungry cat meowing but not barking?)} The dog is excited. \textit{(Is the excited dog barking but not hungry?)} \\
			Many Memories &
			There is a black, meowing cat. There is a large, wooden plane. The cat is hungry. \textit{(Is the hungry cat meowing but not large nor wooden?)} There is a beautiful unknown song. The plane is large. \textit{(Is the large plane wooden, but not a song nor a cat?)} There is a desolate, windy, desert. The song is beautiful. \textit{(Is the beautiful song unknown, but not a desert, nor a cat nor a plane?)} The desert is desolate. \textit{(Is the desolate desert windy, but not a song, nor a cat nor a plane?)} \\
			Problem of Two &
			There is a black meowing cat. There is a small orange cat. \textit{(Is the black cat meowing, but not small nor orange?)} \textit{(Is the orange cat small, but not black nor meowing?)} \\
			\hline
		\end{tabularx}
		\endgroup
		\tablenote{
			Example sentences to illustrate each test.
			These examples are written in English, and while they were constructed to functionally match the artificial sentences used in our simulations, they are not direct translations of them and contain differences in their structure and semantics.
			See \autoref{tab:example_test_specifications} for examples of the artificial sentences used in our simulations.
			We assume that the definite determiner \enquote{the} can be used both for introducing a new referent and for referring back to an earlier mentioned referent, whereas the indefinite constuction \enquote{There is a} always signals the introduction of a new referent.
			In brackets we provide the most important questions we query from the model's responses.
			We performed additional queries of the model's current memory state, but have ommitted these from the table when they were not important for understanding the test.
			When the query contains a \enquote{not} we expect the model to not retrieve the negated attribute.
			Ellipses indicate the passage of time, which is important for testing the model's ability to maintain and forget information over time.
			Note that all discourses are logically consistent, but the sentences are often picked to be semantically uncorrelated so that recall is not reliant on long-term semantic knowledge.
		}\label{tab:example_sentences}
	\end{table*}

	\newcolumntype{Y}{>{\raggedright\arraybackslash}X}
	\begin{table*}
		\caption{Example test specifications (artificial sentences)}
		\small
		\begingroup
		\setlength\tabcolsep{4pt}

		\begin{tabularx}{\textwidth}{>{\centering\arraybackslash}X}
			\hline
			\textbf{Semantic association} (category): Deduced association (test)\\
			\textit{Newborn implies small, allowing deduction that these sentences refer to the same entity (i.e. the cat).} \\
		\end{tabularx} \\
		\begin{tabularx}{\textwidth}{l|XXXX|XXXX}
			\textbf{time $\rightarrow$}	& 0ms			& 500ms						& 1000ms										& 1500ms				& 2000ms				& 2500ms					& 3000ms										& 3500ms				\\
			\hline
			\textbf{artificial}	& \textcolor{blue}{cova}& \textcolor{blue}{covisec}	& \textit{response}								& \textit{eos}			& \textcolor{blue}{avip}& \textcolor{blue}{cososeb}	& \textit{response}								& \textit{eos}			\\
			\textbf{natural}	& newborn				& cat						& 												& . 					& small					& black						&												& .						\\
			\textbf{expected}	& \cellcolor{gray!30}	& \cellcolor{gray!30}		& \cellcolor{green!20} \textcolor{blue}{cova}	&  \cellcolor{gray!30}	& \cellcolor{gray!30}	& \cellcolor{gray!30}		& \cellcolor{green!20} \textcolor{blue}{cova}	& \cellcolor{gray!30}	\\
								& \cellcolor{gray!30}	& \cellcolor{gray!30}		& \cellcolor{green!20} \textcolor{blue}{covisec}&  \cellcolor{gray!30}	& \cellcolor{gray!30}	& \cellcolor{gray!30}		& \cellcolor{green!20} \textcolor{blue}{covisec}& \cellcolor{gray!30}	\\
								& \cellcolor{gray!30}	& \cellcolor{gray!30}		& \cellcolor{green!20} 							&  \cellcolor{gray!30}	& \cellcolor{gray!30}	& \cellcolor{gray!30}		& \cellcolor{green!20} \textcolor{blue}{avip}	& \cellcolor{gray!30}	\\
								& \cellcolor{gray!30}	& \cellcolor{gray!30}		& \cellcolor{green!20} 							&  \cellcolor{gray!30}	& \cellcolor{gray!30}	& \cellcolor{gray!30}		& \cellcolor{green!20} \textcolor{blue}{cososeb}& \cellcolor{gray!30}	\\
			\hline
		\end{tabularx} \\

		\begin{tabularx}{\textwidth}{>{\centering\arraybackslash}X}
			\textbf{Controlled storage} (category): Inferred novelty (test)\\
			\textit{Lack of overlapping attributes allows for inference that second sentence refers to a different entity.} \\
		\end{tabularx} \\
		\begin{tabularx}{\textwidth}{l|XXXX|XXXX}
			\textbf{time $\rightarrow$}	& 0ms						& 500ms						& 1000ms									& 1500ms				& 2000ms					& 2500ms						& 3000ms									& 3500ms				\\
			\hline
			\textbf{artificial}	& \textcolor{blue}{avip}	& \textcolor{blue}{cos}		& \textit{response}							& \textit{eos}			& \textcolor{orange}{cov}	& \textcolor{orange}{vamup}		& \textit{response}							& \textit{eos}			\\
			\textbf{natural}	& small						& pot						& 											& . 					& big						& chair							&											& .						\\
			\textbf{expected}	& \cellcolor{gray!30}		& \cellcolor{gray!30}		& \cellcolor{green!20} \textcolor{blue}{avip}& \cellcolor{gray!30}	& \cellcolor{gray!30}		& \cellcolor{gray!30}			& \cellcolor{green!20} \textcolor{orange}{cov}	& \cellcolor{gray!30}	\\
								& \cellcolor{gray!30}		& \cellcolor{gray!30}		& \cellcolor{green!20} \textcolor{blue}{cos}& \cellcolor{gray!30}	& \cellcolor{gray!30}		& \cellcolor{gray!30}			& \cellcolor{green!20} \textcolor{orange}{vamup}& \cellcolor{gray!30}	\\
			\textbf{not}		& \cellcolor{gray!30}		& \cellcolor{gray!30}		& \cellcolor{gray!30} 						& \cellcolor{gray!30}	& \cellcolor{gray!30}		& \cellcolor{gray!30}			& \cellcolor{red!20} \textcolor{blue}{avip}	& \cellcolor{gray!30}	\\
			\textbf{expected}	& \cellcolor{gray!30}		& \cellcolor{gray!30}		& \cellcolor{gray!30} 						& \cellcolor{gray!30}	& \cellcolor{gray!30}		& \cellcolor{gray!30}			& \cellcolor{red!20} \textcolor{blue}{cos}	& \cellcolor{gray!30}	\\
			\hline
		\end{tabularx} \\

		\begin{tabularx}{\textwidth}{>{\centering\arraybackslash}X}
			\textbf{Memory retrieval} (category): Pattern completion (test)\\
			\textit{Activated attributes that are part of a bundle activate other associated attributes, even if memory is fully suppressed in between.} \\
		\end{tabularx} \\
		\begin{tabularx}{\textwidth}{l|XXXXX|XX|XXXX}
			\textbf{time $\rightarrow$}	& 0ms						& 500ms						& 1000ms						& 1500ms									& 2000ms				& 2500ms									& 3000ms					& 3500ms					& 4000ms									& 4500ms				& 5000ms \\
			\hline
			\textbf{artificial}	& \textcolor{blue}{amuf}	& \textcolor{blue}{avifep}	& \textcolor{blue}{amupoc}	& \textit{response}							& \textit{eos}			& \textit{response}							& \textit{eos}			& \textcolor{blue}{amuf}	& \textcolor{blue}{avifep}	& \textit{response}							& \textit{eos}			\\
			\textbf{natural}	& heavy						& grey						& box						&											& .						&											& .			& heavy						& grey						& 											& .						\\
			\textbf{expected}	& \cellcolor{gray!30}		& \cellcolor{gray!30}		& \cellcolor{gray!30}		& \cellcolor{green!20} \textcolor{blue}{amuf}& \cellcolor{gray!30}	& \cellcolor{gray!30}	& \cellcolor{gray!30}						& \cellcolor{gray!30}		& \cellcolor{gray!30}		& \cellcolor{green!20} \textcolor{blue}{amupoc}& \cellcolor{gray!30}	\\
								& \cellcolor{gray!30}		& \cellcolor{gray!30}		& \cellcolor{gray!30}		& \cellcolor{green!20} \textcolor{blue}{avifep}& \cellcolor{gray!30}& \cellcolor{gray!30}	& \cellcolor{gray!30}						& \cellcolor{gray!30}		& \cellcolor{gray!30}		& \cellcolor{green!20} 						& \cellcolor{gray!30}	\\
								& \cellcolor{gray!30}		& \cellcolor{gray!30}		& \cellcolor{gray!30}		& \cellcolor{green!20} \textcolor{blue}{amupoc}& \cellcolor{gray!30}& \cellcolor{gray!30}	& \cellcolor{gray!30}						& \cellcolor{gray!30}		& \cellcolor{gray!30}		& \cellcolor{green!20} 						& \cellcolor{gray!30}	\\
			\textbf{not}		& \cellcolor{gray!30}		& \cellcolor{gray!30}		& \cellcolor{gray!30}		& \cellcolor{gray!30} 						& \cellcolor{gray!30}	& \cellcolor{red!20} \textcolor{blue}{amupoc} & \cellcolor{gray!30}		& \cellcolor{gray!30}		& \cellcolor{gray!30} 						& \cellcolor{gray!30}	& \cellcolor{gray!30}	\\
			\textbf{expected}	& \cellcolor{gray!30}		& \cellcolor{gray!30}		& \cellcolor{gray!30}		& \cellcolor{gray!30} 						& \cellcolor{gray!30}	& \cellcolor{red!20} \textcolor{blue}{avifep}	& \cellcolor{gray!30}		& \cellcolor{gray!30}		& \cellcolor{gray!30} 						& \cellcolor{gray!30}	& \cellcolor{gray!30}	\\
								& \cellcolor{gray!30}		& \cellcolor{gray!30}		& \cellcolor{gray!30}		& \cellcolor{gray!30} 						& \cellcolor{gray!30}	& \cellcolor{red!20} \textcolor{blue}{amuf} 	& \cellcolor{gray!30}		& \cellcolor{gray!30}		& \cellcolor{gray!30} 						& \cellcolor{gray!30}	& \cellcolor{gray!30}	\\
			\hline
		\end{tabularx} \\
		\endgroup
		\begingroup
		\setlength\tabcolsep{2.75pt}

		\begin{tabularx}{\textwidth}{>{\centering\arraybackslash}X}
			\textbf{Interleaved memories} (category): Problem of two (test)\\
			\textit{Overlapping attribute (big) between two entities requires disambiguation before retrieving the correct bundle.} \\
		\end{tabularx} \\
		\begin{tabular*}{\linewidth}{l|lllll|lllll|llll|llll}
			\textbf{time $\rightarrow$}	& 0s	& 	& 1s	& 	& 2s	& 	& 3s	& 	& 4s	& 	& 5s	& 	& 6s	& 	& 7s	& 	& 8s	& 	\\
			\hline
			\textbf{artificial}	& {\color{blue} amupa}	& {\color{blue} cov}	& {\color{blue} vav}	& \textit{resp}					& \textit{eos}	& {\color{orange} cososec}	& \splitcolor{blue}{orange}{cov}	& {\color{orange} covisec}	& \textit{resp}								& \textit{eos}	& \splitcolor{blue}{orange}{cov}	& {\color{blue} amupa}	& \textit{resp}						& \textit{eos}	& \splitcolor{blue}{orange}{cov}	& {\color{orange} covisec}	& \textit{resp}						& \textit{eos}	\\
			\textbf{natural}	& dull					& big					& ball					& 						& .		& cute						& big								& cat						& 									& .		& big								& dull					& 							& .		& big								& cat						& 							& .		\\
			\textbf{expected}	& \cellcolor{gray!30}	& \cellcolor{gray!30}	& \cellcolor{gray!30}	& \cellcolor{green!20} {\color{blue} cov}	& \cellcolor{gray!30}	& \cellcolor{gray!30}	& \cellcolor{gray!30}	& \cellcolor{gray!30}	& \cellcolor{green!20} \splitcolor{blue}{orange}{cov}	& \cellcolor{gray!30}	& \cellcolor{gray!30}	& \cellcolor{gray!30}	& \cellcolor{green!20} {\color{blue} vav}		& \cellcolor{gray!30}	& \cellcolor{gray!30}	& \cellcolor{gray!30}	& \cellcolor{green!20} {\color{orange} cososec}	& \cellcolor{gray!30}	\\
								& \cellcolor{gray!30}	& \cellcolor{gray!30}	& \cellcolor{gray!30}	& \cellcolor{green!20} {\color{blue} amupa}	& \cellcolor{gray!30}	& \cellcolor{gray!30}	& \cellcolor{gray!30}	& \cellcolor{gray!30}	& \cellcolor{green!20} {\color{orange} cososec}			& \cellcolor{gray!30}	& \cellcolor{gray!30}	& \cellcolor{gray!30}	& \cellcolor{green!20} 							& \cellcolor{gray!30}	& \cellcolor{gray!30}	& \cellcolor{gray!30}	& \cellcolor{green!20}							& \cellcolor{gray!30}	\\
								& \cellcolor{gray!30}	& \cellcolor{gray!30}	& \cellcolor{gray!30}	& \cellcolor{green!20} {\color{blue} vav}	& \cellcolor{gray!30}	& \cellcolor{gray!30}	& \cellcolor{gray!30}	& \cellcolor{gray!30}	& \cellcolor{green!20} {\color{orange} covisec}			& \cellcolor{gray!30}	& \cellcolor{gray!30}	& \cellcolor{gray!30}	& \cellcolor{green!20}							& \cellcolor{gray!30}	& \cellcolor{gray!30}	& \cellcolor{gray!30}	& \cellcolor{green!20}							& \cellcolor{gray!30}	\\
			\textbf{not}		& \cellcolor{gray!30}	& \cellcolor{gray!30}	& \cellcolor{gray!30}	& \cellcolor{gray!30}						& \cellcolor{gray!30}	& \cellcolor{gray!30}	& \cellcolor{gray!30}	& \cellcolor{gray!30}	& \cellcolor{red!20} {\color{blue} amupa}				& \cellcolor{gray!30}	& \cellcolor{gray!30}	& \cellcolor{gray!30}	& \cellcolor{red!20} {\color{orange} cososec}	& \cellcolor{gray!30}	& \cellcolor{gray!30}	& \cellcolor{gray!30}	& \cellcolor{red!20} {\color{blue} amupa}		& \cellcolor{gray!30}	\\
			\textbf{expected}	& \cellcolor{gray!30}	& \cellcolor{gray!30}	& \cellcolor{gray!30}	& \cellcolor{gray!30}						& \cellcolor{gray!30}	& \cellcolor{gray!30}	& \cellcolor{gray!30}	& \cellcolor{gray!30}	& \cellcolor{red!20} {\color{blue} vav}					& \cellcolor{gray!30}	& \cellcolor{gray!30}	& \cellcolor{gray!30}	& \cellcolor{red!20} {\color{orange} covisec}	& \cellcolor{gray!30}	& \cellcolor{gray!30}	& \cellcolor{gray!30}	& \cellcolor{red!20} {\color{blue} vav}			& \cellcolor{gray!30}	\\
			\hline
		\end{tabular*}
		\endgroup
		\tablenote{\small Example test stimuli for each of the test categories. 
			Each example test is shown in the artificial language as provided to the network (top row) and as a reduced English translation (second row).
			In addition, the expected response of the network per word is provided in the rows below, split into positive expectations, i.e. features that should be present, and negative expectations, i.e. features that should be absent.
			Test performance is defined as the similarity of the network's response to this expected response.
			The \textit{resp(onse)} and \textit{eos} symbols indicate to the network that its current state is read out as a response or that the sentence has ended, respectively.
			Gray cells indicate that no specific response is expected at that moment, while green and red cells indicate that certain features are expected to be present or absent, respectively.
			Words shown in the same colour are part of the same bundle.
			This information is not provided to the network, but must be inferred from syntax (e.g. eos) or semantics (e.g. conflicting features).
			Note that the words in the artificial language have semantically arbitrary meanings, and that as such the English translations were selected to match semantic propties, not on the basis of the actual meaning of the words.
			Also note that the artificial language contains no part-of-speech information, meaning that the difference between nouns and adjectives that is present in the English translations is not present in the artificial language itself.
			The closest translation would be to consider all words as adjectives (i.e. properties), e.g. \enquote{cat-like} instead of \enquote{cat}, and to end each phrase with \enquote{entity} -- e.g. \enquote{The newborn cat-like entity}.}\label
			{tab:example_test_specifications}
	\end{table*}

\end{document}